\documentclass{appolb}

\usepackage{amssymb}
\usepackage{hyperref}
\usepackage{graphicx}
\usepackage{amsmath}

\usepackage[numbers,sectionbib,sort&compress]{natbib}

\begin{document}

\title{Transverse densities of the energy-momentum tensor \\ and the gravitational form factors of the pion%
\thanks{Submitted to Acta Physica Polonica B special volume dedicated to
Dmitry Diakonov, Victor Petrov, and Maxim Polyakov.}}

\author{Wojciech Broniowski$^{1,2}$\thanks{Wojciech.Broniowski@ifj.edu.pl} and Enrique Ruiz Arriola$^{3}$\thanks{earriola@ugr.es} 
\address{$^{1}$The H. Niewodnicza\'nski Institute of Nuclear Physics, Polish Academy \\ of Sciences, 31-342~Cracow, Poland}
\address{$^{2}$Institute of Physics, Jan Kochanowski University, 25-406~Kielce, Poland}
\address{$^{3}$Departamento de F\'{\i}sica At\'{o}mica, Molecular y Nuclear and Instituto Carlos I \\ de  F{\'\i}sica Te\'orica y Computacional}}

\maketitle

\begin{abstract}
We present general features of the transverse densities of the
stress-energy-momentum tensor $\Theta^{\mu\nu}$ in the pion. We 
show positivity of the transverse density of $\Theta^{++}$ (analogous to the positivity of the 
transverse density of the electromagnetic current $J^+$) and 
discuss its consequences 
in conjunction with analyticity and quark-hadron duality, as well as the
connection to $\pi\pi$ scattering at low energies. Our analysis
takes into account the perturbative QCD effects, dominating at high
momenta (or low transverse coordinate $b$), the effects of Chiral
Perturbation Theory, dominating at low momenta (high $b$), and meson dominance in
the intermediate region.  We incorporate constraints form analyticity,
leading to sum rules for the spectral densities of the corresponding
form factors, which {\em i.a.} are relevant for the high-momentum (or the low-$b$) asymptotics.
With the obtained high- and low-$b$ behavior, we deduce that the scalar (trace-anomaly) gravitational transverse density 
 $\Theta^{\mu}_\mu(b)$ must change sign, unlike the case of the positive definite $J^+(b)$ or $\Theta^{++}(b)$.
We also discuss the transverse pressure in the pion, which is positive and singular at low $b$, and negative at high $b$, in harmony with 
the stability criterion. The results for the form factors for space-like 
momenta are compared to the recent lattice QCD data. 
\end{abstract}

\bigskip
  
 \section{Remembrance \label{sec:rem}}

The horn signal used to call everyone for coffee to the window-less
social room of the Theoretical Physics Department II of the Ruhr
Universit\"at in Bochum. There, among all the friends of the group
headed by dearest Klaus Goeke, many brilliant ideas were coined and
discussed, with the three friends commemorated in this volume playing
leading roles.  Out of the many concepts that originated there, those
from Maxim and collaborators continue to be particularly important in
our humble research.  In particular, the $D$ (druck)
term~\cite{Polyakov:1999gs} in the generalized parton distributions,
hence in the gravitational form factors (GFFs) related to matrix
elements of the stress-energy momentum (SEM) tensor, as well as the
interpretation of the matrix elements of the energy-momentum tensor
via the physically intuitive mechanistic properties of
hadrons~\cite{Polyakov:2002yz,Polyakov:2018zvc}, are at the core of in
this paper.

\section{Stress-energy-momentum tensor in the pion \label{sec:empion}}

\subsection{Introduction and scope}

The SEM tensor appears as a universal Noether current in any
Quantum Field Theory textbook since the early days.  It has played a
decisive role in the theoretical understanding of scale invariance and its violations~\cite{Carruthers:1971vz}. Yet, phenomenological implications
for hadronic physics have been less frequent, mainly due to the
impossibility of making direct experimental or {\it ab initio}
determinations.
The very first mention of the gravitational form factors that we are
aware of was within the context of $NN$ interactions, in order to
characterize the tensor meson
exchange~\cite{sharp1963asymptotic}. Pagels made the first study based
on analyticity and final state interactions~\cite{Pagels:1966zza}.  The proper
tensor decomposition was written down by Raman~\cite{Raman:1971jg}.
Coupled channel analyses have been pioneered in
Refs.~\cite{Truong:1989my,Gasser:1990bv}, whereas a Chiral Perturbation
Theory ($\chi$PT) setup was proposed in~\cite{Donoghue:1991qv}.

The recent activity in GFFs of the pion has been largely spurred by
the release of accurate lattice QCD data by the MIT
group~\cite{Hackett:2023nkr,Pefkou:2023okb}, obtained for all the
parton species and close to the physical point, with
$m_\pi=170$~MeV. This ameliorates the seminal studies of the quark
parts~\cite{Brommel:2007zz,QCDSF:2007ifr}, recently repeated for
$m_\pi = 250$~MeV~\cite{Delmar:2024vxn}, of the gluonic
parts~\cite{Shanahan:2018pib}, or the trace anomaly
component~\cite{Wang:2024lrm} at large $m_\pi$. On the experimental
side, a method of extracting GFFs from the $\gamma \gamma^\ast \to
\pi^0 \pi^0$ data~\cite{Belle:2015oin} was developed
in~\cite{Kumano:2017lhr}, with further prospects at Super-KEKB and
ILC, where the generalized distribution amplitudes of the pion could
be investigated.

Model calculations of the pion GFFs have been carried out in various
approaches,
including~\cite{Broniowski:2007si,Broniowski:2008hx,Frederico:2009fk,%
  Masjuan:2012sk,Fanelli:2016aqc,Freese:2019bhb,Krutov:2020ewr,Xing:2022mvk,Xu:2023izo,Li:2023izn,Liu:2024jno,Liu:2024vkj,%
  Wang:2024sqg,Sultan:2024hep,Fujii:2024rqd,Krutov:2024adh}.  In this
paper we extend the meson dominance
model~\cite{Broniowski:2024oyk,RuizArriola:2024uub}, applied
successfully to the lattice data
of~\cite{Hackett:2023nkr,Pefkou:2023okb} (our techique was later repeated in~\cite{Cao:2024zlf}), by incorporating explicitly
the perturbative QCD (pQCD), computed recently~\cite{Tong:2021ctu,Tong:2022zax}, as well as the  $\chi$PT~\cite{Donoghue:1991qv}
pieces. Our analysis is done with the help of the dispersion
relations, which allows us to preserve analyticity when combining
various contribution (meson dominance, pQCD, $\chi$PT) and to satisfy
all the required sum rules.

We obtain the transverse densities of the energy-momentum
tensor in the pion and discuss their general features, such as the
singular limits at a low transverse coordinate $b$, stemming from
pQCD, and the high-$b$ asymptotics following from the threshold
behavior of the $\pi\!-\!\pi$ scattering. For the scalar (trace anomaly)
transverse density, the opposite sign of the low- and high-$b$ limits
proves that it has to change sign as a function of $b$. This is not
the case of the charge or the tensor GFFs, which are positive definite.
Finally, we discuss the transverse pressure inside the pion, following
from the gravitational transverse densities.

\subsection{The energy momentum tensor}

We will use the Hilbert definition of the SEM tensor via the coupling
to gravity, which in the QCD case coincides with the
Belinfante-Rosenfeld definition~\cite{Pokorski:1987ed} (see also
Appendix E of~\cite{Belitsky:2005qn}),
\begin{eqnarray}
\hspace{-2mm}   \Theta^{\mu\nu} = \tfrac{i}4 \bar \Psi \left[ \gamma^\mu  \overleftrightarrow{D}^\mu \!+\!
    \gamma^\nu  \overleftrightarrow{D}^\mu \right] \Psi - F^{\mu \lambda a} F^\nu_{\lambda
    a} + \tfrac14 g^{\mu \nu} F^{\sigma \lambda a} F_{\sigma \lambda a} + \Theta_{\rm GF-EOM}^{\mu\nu},
\end{eqnarray}  
where in the quantized case one has in addition the gauge-fixing and the
equations-of-motion terms. This object is local, symmetric,
$\Theta^{\mu\nu}=\Theta^{\nu\mu}$, conserved $\partial_\mu
\Theta^{\mu\nu}=0$, and from the Lorentz group can be irreducibly
decomposed as a sum of a traceful and traceless parts. It is a highly singular operator which requires
renormalization. The trace can be written as an anomalous
divergence of the dilatation current, $D^\mu= x_\nu \Theta^{\mu\nu}$,
related unambiguously to the scale invariance breaking~\cite{Pokorski:1987ed}, namely
\begin{eqnarray}
\Theta \equiv \Theta^\mu_\mu=\frac{\beta(\alpha)}{4\alpha} {G^{\mu \nu}}^2+ [1+\gamma_m(\alpha)] \sum_f m_f \bar{q}_f q_f. \label{eq:anom}
\end{eqnarray} 
Here $\beta(\alpha) = \mu^2 d \alpha / d \mu^2 = - \alpha [\beta_0 (\alpha/4\pi) +
  {\cal O}(\alpha^2)  ] < 0 $ is the QCD beta function with $\beta_0 = (11 N_c- 2N_F)/3$, 
  $N_c$ is the number of colors, $N_F$ is the number of flavors, 
$\gamma_m(\alpha)= 2 \alpha/\pi +{\cal O}(\alpha^2) $ is the quark mass anomalous dimension, and  $f$ enumerates active flavors. One has 
$\alpha(t) = (4\pi /\beta_0) / \ln \left (-t/\Lambda_{\rm QCD}^2 \right)
$ with $\alpha(t) $ real for $t=-Q^2 <0$. We take $\Lambda_Q=225$~MeV and $N_F=3$ in our numerical studies presented later on.

\subsection{Raman decomposition}

The standard tensor decomposition of the SEM tensor matrix element in the pion state of isospin $a,b$ is 
\begin{eqnarray}
&& \hspace{-7mm}  \langle \pi^a (p') | \Theta^{\mu \nu}(0) | \pi^b (p) \rangle = \delta^{ab} \left [ 2 P^\mu P^\nu A(q^2)  + \frac12 ( q^\mu q^\nu - g^{\mu \nu} q^2 ) D(q^2 ) \right ], \label{eq:tmunudef}
\end{eqnarray}
where $P=\tfrac{1}{2}(p'+p)$, $q=p'-p$, and $A$ and $D$ are the gravitational form factors. For the electromagnetic current, $J_Q^\mu = \bar \Psi \gamma^\mu Q \Psi $, one has
\begin{eqnarray}
\langle \pi^\pm (p') |J_Q^\mu(0) | \pi^\pm (p) \rangle  =  \pm 2 P^\mu F(q^2),  \label{eq:Jmudef}
\end{eqnarray}
where $F$ is the charge form factor, to which we shall make frequent
references in the context of our discussion of GFFs for comparison purposes.

The proper decomposition into a sum of two separately conserved irreducible tensors of a
well-defined total angular momentum,  $J^{PC}=0^{++}$ (scalar) and $2^{++}$ (tensor), has the form~\cite{Raman:1971jg}
\begin{eqnarray}
&& \Theta^{\mu \nu} =  \Theta_S^{\mu \nu}+\Theta_T^{\mu \nu},  \;\;  \Theta_S^{\mu \nu} = \frac13 Q^{\mu \nu} \Theta, \\ 
&&  \Theta_T^{\mu \nu} = \Theta^{\mu \nu}- \frac13  Q^{\mu \nu} \Theta  = 2 \left[ P^\mu P^\nu - \frac{P^2}3  Q^{\mu \nu} \right] A, \nonumber
\end{eqnarray} 
where $Q^{\mu \nu}\equiv g^{\mu \nu}-{q^\mu q^\nu}/{q^2}$ (for brevity, we drop here the argument of the GFFs, which is $q^2$). This
decomposition implements a separate conservation of the scalar and
tensor parts, $ q_\mu \Theta_S^{\mu\nu}=q_\mu
\Theta_T^{\mu\nu}= 0$, unlike the and often used  naive decomposition
with $\Theta_T^{\mu \nu} = \Theta^{\mu\nu}- \tfrac14 g^{\mu\nu} \Theta  $ and
$\Theta_S^{\mu\nu}=\tfrac14 g^{\mu\nu} \Theta$, which violated the separate
conservation property, since, e.g.,  $q_\mu \Theta_S^{\mu\nu}= q^\nu \Theta $. 
The Raman separation of
the GFFs, when promoted to the operator level in QCD, also has the
feature that the trace of SEM is singular and requires
renormalization, cf. Eq.~(\ref{eq:anom}). From this point of view, $\Theta$
and $A$, carrying good $J^{PC}$ quantum numbers, should be regarded as the basic form
factors~\cite{Broniowski:2024oyk}, whereas $D$ mixes the quantum numbers and is given as a
secondary quantity (albeit naturally appearing in the mechanistic properties~\cite{Polyakov:2002yz,Polyakov:2018zvc}) by the combination
\begin{eqnarray}
D= -\frac{2}{3q^2} \left [ \Theta - \left ( 2 m_\pi^2 -\tfrac{1}{2}\, q^2 \right ) A \right]. \label{eq:Drel}
\end{eqnarray}

\subsection{Meson dominance vs lattice data}

At a phenomenological level, we have found that from the MIT lattice data~\cite{Hackett:2023nkr} at
$m_\pi=170$~MeV  one can infer, using $\chi$PT to NLO, that in the range 
$ 0<-t < 2~{\rm GeV}^2$ and for the physical $m_\pi=140$~MeV the GFFs can be very satisfactorily described in the single resonance
saturation picture proposed long ago~\cite{Donoghue:1991qv},
\begin{eqnarray}
  \Theta(t) &=& 2 m_\pi^2  + \frac{t m_\sigma^2}{m_\sigma^2-t} \\
  A(t) &=& \frac{m_{f_2}^2}{m_{f_2}^2-t},
\label{eq:dom}
\end{eqnarray}
with $m_\sigma =0.63(6)$~GeV  and $m_{f_2}=1.27(4)$~GeV~\cite{Broniowski:2024oyk}.
The gravitational low energy constants $L_{11,12,13}$, as well as
the corresponding value of the $D$-term, $ D(0) = -0.95(3)$, have been extracted thereof~\cite{RuizArriola:2024uub}.

\section{Transverse densities \label{sec:tran}}

\subsection{Preliminary}

The transverse densities of hadrons in the infinite momentum frame were advocated
in~\cite{PhysRevD.15.1141,Burkardt:2000za,Diehl:2002he,Burkardt:2002hr,Miller:2010nz,Freese:2022fat}
as proper objects relating to the probabilistic interpretation of the
parton distributions. In particular, the transverse electromagnetic
(charge) density can be shown to be manifestly
positive definite~\cite{Burkardt:2002hr,Pobylitsa:2002iu,Diehl:2002he}. Likewise, the transverse
$\Theta^{++}$ distribution is also positive definite (for any hadronic state), as we will demonstrate below.  Moreover, as
argued in~\cite{Freese:2021mzg}, for spin-0 mesons such as the pion,
the 3D Breit-frame densities are not related to the transverse
densities via the Abel transform, as is the case for the nucleon~\cite{Panteleeva:2021iip}, hence they acquire even more significance.

\subsection{Light-cone kinematics for plane waves}

We take the conventions $p^\pm = (p^0
\pm p^3)/\sqrt{2}= p_\mp $, such that $x\cdot p= p^+ x^- + p^- x^+ - p_\perp
\cdot x_\perp $ and $d^4 p = dp^+ dp^- d^2 p_\perp $. Also,
$g^{++}=g^{--}=0 $ and $g^{+-}=g^{-+}=1$.
Assume the momentum transfer has only the $q_\perp$ component, i.e. 
$q^+=q^-=0$. 
The momenta of the pions in the chosen frame (IMBF, infinite momentum Breit frame) are therefore
\begin{eqnarray}
p^+={p^+}'=P^+, \;\; p^-={p^-}'=\frac{\tfrac{1}{4}q_\perp^2+m_\pi^2}{2{P^+}}, \;\;p_\perp=-{p_\perp}'=-\tfrac{1}{2}q_\perp.  \label{eq:kinem}
\end{eqnarray}
In the light front quantization (see, {\em e.g.},~\cite{Freese:2022fat} and references therein),  
the states on the mass shell are labeled as $|p^+,p_\perp\rangle$. They fulfill the completeness relation 
\begin{eqnarray} 
\int \frac{dp^+ dp^- d^2p_\perp}{(2\pi)^4} 2\pi \theta(p^+)\, \delta(2 p^+ p^- -p_\perp^2-m_\pi^2) | p^+, p_\perp  \rangle  \langle p^+ , p_\perp |= 1, 
\end{eqnarray}
corresponding to  $p^- = (p_\perp^2+m_\pi^2)/2p^+$ and $p^+ >0$,
and the invariant normalization 
\begin{eqnarray}
\langle p^+ , p_\perp | {p^+}', {p_\perp}'  \rangle  = 2p^+ (2\pi)^3 \delta({p^+} - {p^+}') \delta^2(p_\perp-{p_\perp}'), \label{eq:lfn}
\end{eqnarray}
which implies the rule to divide with $\sqrt{2p^+}$ for the initial and $\sqrt{{2p^+}'}$ for the final state.

The matrix element of the $+$ component of the electromagnetic current is 
$2p^+ F(q_\perp^2)$, and the factor of $2p^+$ cancels with the one originating 
from the normalization~(\ref{eq:lfn}) (here this is similar to the instant form case). One defines the transverse charge density as
\begin{eqnarray}
F^+(b) = \int \frac{d^2q_\perp}{ (2\pi)^2} e^{-i q_\perp \cdot b} F(q_\perp^2). \label{eq:F1}
\end{eqnarray}
Note that $F^+(b)$ corresponds to the $+$ component of the current. The $\perp$ components of the current vanish identically in IMBF.
For the $-$ component of the current, we acquire the factor $P^-/P^+$ in the integrand, 
which vanishes in IMBF when $P^+\to \infty$.  

Next, let us  look at the $++$ component of Eq.~(\ref{eq:tmunudef}),
\begin{eqnarray}
\Theta^{++}(b) = \int \frac{d^2q_\perp}{2P^+  (2\pi)^2} e^{-i q_\perp \cdot b}  \, 2{P^+}^2 A(q_\perp^2) = P^+ A(b), \label{eq:Th1}
\end{eqnarray}
which gives a simple interpretation to $A(b)$ as the relative distribution of $P^+$ in the transverse coordinate space. 
Obviously, $\int d^2b \,\Theta^{++}(b) = P^+$.
For the $+-$ component  we have 
\begin{eqnarray}
\Theta^{+-}(b) = \int \frac{d^2q_\perp}{2P^+  (2\pi)^2} e^{-i q_\perp \cdot b}  \left [2{P^+}P^- A(q_\perp^2) + \frac{1}{2}q_\perp^2 D(q_\perp^2)\right ],
\end{eqnarray}
where we have used $g^{+-}=1$. With the kinematics~(\ref{eq:kinem}) we get immediately
\begin{eqnarray}
\Theta^{+-}(b) = \frac{1}{2 P^+} \int \frac{d^2q_\perp}{ (2\pi)^2} e^{-i q_\perp \cdot b}  
\left [ (m_\pi^2+\tfrac{1}{4}q_\perp^2) A(q_\perp^2) + \frac{1}{2}q_\perp^2 D(q_\perp^2)\right ].
\end{eqnarray}
The $--$ component 
is strongly suppressed, $\sim 1/P_+^3$, and involves only $A$ in the chosen frame. The transverse components are 
\begin{eqnarray}
&& \Theta^{ij}(b) =  \frac{1}{2 P^+} \int \frac{d^2q_\perp}{(2\pi)^2} e^{-i q_\perp \cdot b}  \frac{1}{2}  \left [ q_\perp ^i q_\perp^j  - \delta^{ij} q_\perp^2 \right ] D(q_\perp^2) = \nonumber \\
&&~~~~~\delta^{ij} p(b)+\left [ \frac{b^i b^j}{b^2}-\frac{1}{2}\delta^{ij}\right ]  s(b), \label{eq:press}
\end{eqnarray}
where $p(b)$ is the transverse pressure and $s(b)$ denotes the transverse shear forces. The trace GFF is 
\begin{eqnarray}
&& \Theta^\mu_\mu(b) = 2\Theta^{+-}(b)-\Theta^{11}(b)-\Theta^{22}(b) = \nonumber \\
&&~~~~~~~ \frac{1}{2P^+} \int \frac{d^2q_\perp}{ (2\pi)^2} e^{-i q_\perp \cdot b} 
\left [ 2(m_\pi^2+\tfrac{1}{4}q_\perp^2) A(q_\perp^2) + \frac{3}{2}q_\perp^2 D(q_\perp^2)\right ] \nonumber \\
&&~~=  \frac{1}{2P^+} \int \frac{d^2q_\perp}{ (2\pi)^2} e^{-i q_\perp \cdot b} \Theta(q_\perp^2) =  \frac{1}{2P^+} \Theta(b).
\end{eqnarray}
The normalizations are $\int d^2 b \, \Theta^\mu_\mu (b) = m_\pi^2/P^+$ and 
$\int d^2 b \, \Theta (b) = 2m_\pi^2$.
In the above formulas the normalization factors of $1/(2P^+)$ factor out of the Fourier 
transforms, in contrast to the instant form quantization, where $1/(2P^0)$ = $1/(2\sqrt{m_\pi^2+q_\perp^2/4})$ remains inside the integral 
and largely affects the interpretation, in particular for light hadrons.

\subsection{Intrinsic properties vs form factors}

From a Quantum Mechanical point of view, intrinsic physical properties
are obtained as expectation values of self-adjoint operators in a
normalizable quantum mechanical state $|\psi \rangle$, $ \langle A
\rangle_\psi = \langle \psi | A | \psi \rangle $. However, form factors
by themselves are non-diagonal matrix elements between plane waves.
The use of the light-cone (LC) coordinates has been promoted as a way to provide a
proper definition of intrinsic properties related to form factors, fully compatible with the
probabilistic interpretation. 
To establish the connection, instead of the plane waves one considers wave packets, which in
LC coordinates and for the pion read
\begin{eqnarray}
  |\phi \rangle = \int \frac{d^2 p_\perp dp^+}{(2\pi)^3 2 p^+}
 \tilde\phi (p_\perp , p^+ ) |p_\perp , p^+ \rangle .
\end{eqnarray}
From here  we have the scalar product 
\begin{eqnarray}
\langle \phi |\psi \rangle &=& \int \frac{d^2 p_\perp dp^+}{(2\pi)^3 2 p^+}
\tilde\phi ( p_\perp , p^+ )^* \tilde\psi (p_\perp, p^+)  \nonumber \\ &=&
\int d^2 x_\perp dx^- \phi ( x_\perp , x^- )^* \psi (x_\perp, x^-).
\end{eqnarray}
The coordinate and momentum representations are related via the Fourier transform,
\begin{eqnarray}
  \psi(x_\perp ,x^-  ) = \int \frac{d^2 p_\perp dp^+}{\sqrt{(2\pi)^3 2 p^+}} \tilde\psi (p_\perp , p^+ ) e^{i (x_\perp \cdot p_\perp - p^+ x^-)}.
\end{eqnarray}
In~\cite{Burkardt:2000za} wave packets localized sharply around $p_z \to \infty$  were considered. An equivalent way
is to integrate over the $x^-$ coordinate in the local operator, and define 
the transverse wave packet distribution in the transverse coordinate $b=x_\perp$,
\begin{eqnarray}
  n_\psi (b) = \int dx^- |\psi(b,x^-)|^2 = \int_0^\infty  \frac{dp^+}{4 \pi p^+ } \left| \int \frac{d^2 p_\perp}{(2\pi)^2}
  e^{i b \cdot p_\perp} \tilde\psi(p_\perp , p^+) \right|^2.
\end{eqnarray}
For local operators, we consider the $x^+=0 $ quantization surface.
Using translational invariance, $O(x)= e^{i P \cdot x} O(0) e^{-i P \cdot x}$, and after some straightforward manipulations, one obtains the intuitive formula for the expectation value of the electromagnetic current $J^\mu$,
\begin{eqnarray}
  \langle \psi | \int dx^- J^{+} (b,x^-) | \psi \rangle &=&      \int d^2 b' n_\psi(b-b') F(b'),
  \label{eq:Jp}
\end{eqnarray}
where $F(b)$ is the Fourier transform of the charge form factor in the space-like momentum space,
\begin{eqnarray}
F(b)=\int \frac{d^2 q_\perp}{(2\pi)^2}  F (-q_\perp^2) e^{-i q_\perp \cdot b }.
\end{eqnarray}

Next, we define
\begin{eqnarray}
  n^+_\psi (b) &=& \int dx^- \psi^* (b,x^-) i \partial^+ \psi (b,x^-)\nonumber \\
  &=& \int_0^\infty  \frac{dp^+}{4 \pi p^+ } p^+ \left| \int \frac{d^2 p_\perp}{(2\pi)^2}
  e^{i b \cdot p_\perp} \tilde\psi(p_\perp , p^+) \right|^2 
\end{eqnarray}  
to obtain, for the quark part  $\Theta_q^{++}$,
\begin{eqnarray}
  \langle \psi | \int dx^- \Theta_q^{++} (b,x^-) | \psi \rangle &=&  \int d^2 b' n_\psi^+  (b-b') A_q(b') ,
  \label{eq:thetapp}
\end{eqnarray}
where
\begin{eqnarray}
A_q(b) = \int \frac{d^2 q_\perp}{(2\pi)^2}  A_q(-q_\perp^2) e^{-i q_\perp \cdot b }.
\end{eqnarray}
Obviously, for a localized wave packet $n_\psi(b) \to \delta^{(2)}(b)$ and $n^+_\psi(b) \to p^+ \delta^{(2)}(b)$, hence one has
\begin{eqnarray}
\hspace{-7mm} \langle \psi | \int dx^- J^+ (b,x^-) | \psi \rangle \to &F(b) ,   \;\;\; \langle \psi | \int dx^- \Theta^{++} (b,x^- ) | \psi \rangle = P^+ A(b), 
\end{eqnarray}
in accordance with Eqs.~(\ref{eq:F1}) and (\ref{eq:Th1}).

\subsection{Positivity \label{sec:pos}}

In QCD, the EM current and SEM in LC
coordinates and with the gauge $A^+=0$ (which is ghost free), one has
\begin{eqnarray}
J^+ &=& \Psi_+^\dagger Q \Psi_+ , \nonumber \\ 
\Theta_q^{++} &=& \frac{i}{2}  \left( \Psi^\dagger_+ \partial^+ \Psi_+ - \partial^+ \Psi_+^\dagger \Psi_+ \right), \;\;\; 
\Theta_g^{++} = (\partial^+ A^a_\perp )^2 , \nonumber \\
\Theta^{++} &=& \Theta_q^{++} + \Theta_g^{++}.
\end{eqnarray}
Here $Q$ is the electric charge,
$\Psi_\pm = {\cal P}_\pm \Psi$, where ${\cal P}_\pm =\gamma^0 \gamma^\pm $ 
are orthogonal projection operators satisfying ${\cal P}_++{\cal P}_-=1$, ${\cal P}_{\pm}^2 = {\cal P}_\pm =
{\cal P}_\pm^\dagger $ and ${\cal P}_\pm {\cal P}_\mp =0$, and the $A^{a\mu}$ is the gluon field.
We note that the gluon component, $\Theta_g^{++}$, is manifestly positive definite. Importantly, 
the sum of the quark and gluon parts is renormalization group 
invariant. 

The field expansion for the quark field in the 
transverse coordinate space~\cite{Diehl:2002he} at $x^+=0$ is
\begin{eqnarray}
  q_+ (b,x^-) =  \int_0^\infty \frac{dp^+}{4\pi p^+} \sum_\lambda && [ b_\lambda (b,p^+) u_{\lambda,+} (p^+) e^{-i p^+ x^-}  \nonumber \\
    &&+ d_\lambda^\dagger (b,p^+) v_{\lambda,+} (p^+) e^{i p^+ x^-} ] ,
\end{eqnarray}
with $b_\lambda^\dagger (b,p^+) $  and $d_\lambda^\dagger (b,p^+) $ denoting 
the particle and antiparticle creation  operators with LC helicity $\lambda$, respectively.  Then
\begin{eqnarray}
  \int d x^- q_+^+ q_+ &=& \sum_\lambda \int \frac{dp^+}{4 \pi p^+} 
  \left[ n (b,p^+) - \bar n_\lambda (b,p^+) \right], \nonumber \\
  \int d x^- q_+^+ i \partial^+ q_+ &=& \sum_\lambda \int \frac{dp^+}{4 \pi p^+} 
 \left[ p^+ n_\lambda (b,p^+) - p^+ \bar n_\lambda (b,p^+)  \right] ,
\end{eqnarray}
with  $n_\lambda (b,p^+)  = b^\dagger_\lambda (b,p^+) b_\lambda (b,p^+) $
and $ \bar n_\lambda (b,p^+)  = d^\dagger_\lambda (b,p^+) d_\lambda (b,p^+) $ denoting
the particle and antiparticle number operators,  respectively. 
Thus, for  $\pi^+ = u \bar d$,
\begin{eqnarray}
  \int dx^-J^+ (b,x^-) 
\underbrace{\to}_{\pi^+} 
 \sum_\lambda \int \frac{dp^+}{4 \pi p^+} 
  \left[  \frac23  n_{u,\lambda} (b,p^+)  + \frac13  n_{\bar d,\lambda} (b,p^+) \right], \label{eq:JJ}
\end{eqnarray}
since generally $q_+^\dagger q_+ $ is positive for quarks and negative for antiquarks. 
Thus~(\ref{eq:JJ}), and consequently $F(b)$, are positive definite.

For $\Theta_q^{++}$
one also finds positivity,
\begin{eqnarray}
&& \hspace{-2cm} \int dx^- \frac{i}{2}  \left(  \Psi^\dagger_+ \partial^+ \Psi_+ -\partial^+ \Psi_+^\dagger
  \Psi_+  \right) = 
  i \int dx^-   \Psi^\dagger_+ \partial^+ \Psi_+  \nonumber  \\
 \hspace{1cm}&\underbrace{\to}_{\pi^+} &
\sum_\lambda \int \frac{dp^+}{4 \pi p^+} 
  \left[  p^+ n_{u,\lambda} (b,p^+)  + p^+  n_{\bar d,\lambda} (b,p^+) \right]
\end{eqnarray}

To summarize this Section, 
both $F(b)$ and $A(b)$ are {\it intrinsic} properties of the pion
which are {\it positive definite}. This allows to interpret $F(b)$ as the
{\it transverse charge distribution}, whereas $A(b)$ is the {\it transverse $p^+ $ 
 distribution} in the pion.  Note that these two distributions do not
contain an interacting piece.  One should keep in mind  that the  formulas and their
interpretation as a whole depend crucially on the LC kinematics and the
gauge $A^+=0$.
We will show below that the parton-hadron duality makes the positivity conditions non-trivial, 
having important implications for the $\pi\pi$ scattering in the elastic region, which {\it must be attractive} 
in the corresponding $J^{PC}$ channels.

In each of the considered channels, $f=F, A, \Theta$, the transverse densities are defined as the Fourier-Bessel transforms
\begin{eqnarray}
f(b) = \int \frac{d^2q_\perp}{ (2\pi)^2} e^{-i q_\perp \cdot b} f(q_\perp^2) = 
\int \frac{q_\perp dq_\perp}{ 2\pi} J_0(b q_\perp) f(q_\perp^2), \label{eq:tranden}
\end{eqnarray}
where $J_0(z)$ is a Bessel function.\footnote{Note that the positivity of  $F(b)$ or $A(b)$ does not necessarily mean
the positivity of the inverse Fourier-transforms into the space-like form factors $F(-Q^2)$  or $A(-Q^2)$.}

\section{Dispersion relations and sum rules \label{sec:drs}}

\subsection{General properties}

All the considered form factors vanish sufficiently fast at large space-like momenta, 
hence satisfy the unsubtracted dispersion relations
\begin{eqnarray}
f(-Q^2)=\frac{1}{\pi} \int_{4m_\pi^2}^\infty ds \frac{{\rm Im}f(s)}{s+Q^2}, \label{eq:disp}
\end{eqnarray}
where $Q^2=-q^2=-t$ is the space-like momentum transfer squared.
The relations make sense, as asymptotically ${\rm Im}\, f(s)$ tends to zero sufficiently fast (see the following) and the integrals converge.
For  $f=F$ or $f=A$, one has the conditions
$f(0)=1$ and $\lim_{Q^2 \to \infty} Q^2 f(Q^2)=0$, which are {\em equivalent} to the two sum rules
\begin{eqnarray}
\frac{1}{\pi} \int_{4m_\pi^2}^\infty ds \frac{{\rm Im}f(s)}{s}=1, \label{eq:csr} \\
\frac{1}{\pi}  \int_{4m_\pi^2}^\infty ds \, {\rm Im}f(s)=0 \label{eq:asr}.
\end{eqnarray}
For the scalar part of the gravitational form factor, since $\Theta(0)=2m_\pi^2$, we get the mass sum rule
\begin{eqnarray}
\frac{1}{\pi} \int_{4m_\pi^2}^\infty ds \frac{{\rm Im}\Theta(s)}{s}=2m_\pi^2. \label{eq:thsr} 
\end{eqnarray}
The once-subtracted form of (\ref{eq:disp}) is 
\begin{eqnarray}
\Theta(-Q^2)=2m_\pi^2-\frac{1}{\pi} \int_{4m_\pi^2}^\infty ds \frac{Q^2}{s}\frac{{\rm Im}\Theta(s)}{s+Q^2}, \label{eq:thdrs} 
\end{eqnarray}
where one can immediately see that the condition 
$\lim_{Q^2 \to \infty} \Theta(-Q^2)=0$ is equivalent to Eq.~(\ref{eq:thsr}). By a miracle of analyticity, the low energy 
condition for $\Theta(0)$ translates into the asymptotic vanishing of $\Theta(-Q^2)$. 
For $\Theta$, there is no sum rule of the form of~(\ref{eq:asr}), since the spectral strength decays too slowly, 
${\rm Im}\Theta(s)\sim 1/ \ln^3 s$ (see Sec.~\ref{sec:consist}).

The slope sum rules,
\begin{eqnarray}
\left . \frac{df(t)}{dt} \right |_{t=0} \equiv f'(0)=  \frac{1}{\pi} \int_{4m_\pi^2}^\infty ds \frac{{\rm Im}f(s)}{s^2} \label{eq:slg}
\end{eqnarray}
have significance when comparing to $\chi$PT and the data. In particular,
\begin{eqnarray}
\Theta'(0)  = \frac{1}{\pi} \int_{4m_\pi^2}^\infty ds \frac{{\rm Im}\Theta(s)}{s^2} = 1 +{\cal O}(m_\pi^2/f_\pi^2), \label{eq:slopeth}
\end{eqnarray}
where $f_\pi=93$~MeV (95~MeV for the lattice $m_\pi=170$~MeV \cite{Broniowski:2024oyk}). 
Combining Eqs.~(\ref{eq:thsr}) and (\ref{eq:slg}) for $\Theta$ we get 
\begin{eqnarray}
2 m_\pi^2 (1\!-\!2 \Theta'(0))= \frac1{\pi} \int_{4 m_\pi^2}^\infty \!\!\! ds  (s-4m_\pi^2)  \frac{{\rm Im} \,\Theta(s)}{s^2}, 
\end{eqnarray}
The left-hand side is negative because from $\chi$PT it follows that $ \Theta'(0)= 1 + {\cal O} (m_\pi^2/f_\pi^2) >1/2$, 
hence there must be a region in $s$ where 
${\rm Im} \,\Theta(s)$ is negative. Since it is positive at the origin, it must change sign.
Sum rule~(\ref{eq:asr}) immediately yields that the spectral strengths of $F$ and $A$ also need to change sign.

\begin{figure}[tb]
\begin{center}
\includegraphics[width=.95\textwidth]{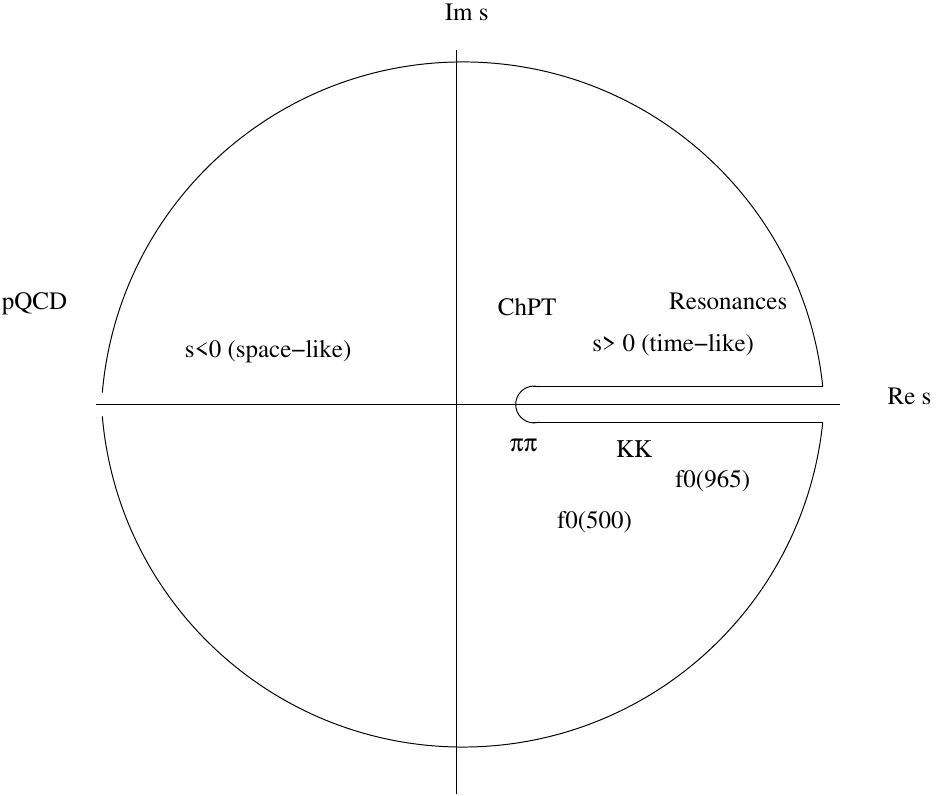}
\caption{Different regions in the complex $s$ plane used in the spectral modeling.}
\label{fig:complex}
\end{center}
\end{figure}

\subsection{Sum rules and modeling}

In modeling form factors in $Q^2$, one usually distinguishes three
regions indicated in Fig.~\ref{fig:complex}: low $Q^2$ dominated by a threshold expansions and/or
$\chi$PT, the high $Q^2$ where pQCD can be applied, and the
intermediate region dominated resonances. The difficulty of modeling
in $Q^2$ lies in appropriate smooth matching of various regions in a
way that preserves analyticity. This is a serious obstacle, as any
step functions or their smoothed versions would unavoidably lead to
spurious behavior in the complex $t$ plane. Moreover, asymptotic
expansions in space-like momenta are obviously not meant {\em per se}
as analytic functions. An alternative, vastly used to avoid these
problems, is to carry out the modeling in the time-like region, which
amounts to assuming proper physically motivated formulas for ${\rm
  Im}\, f(s)$ along the cut. Here also we have three regions: the
low-$s$ range, described with $\chi$PT and extending from the
threshold up to $\Lambda_\chi^2 \sim m_\rho^2$, the high-$s$ range
controlled by pQCD and extending from $\Lambda_p^2$ upwards, and the
intermediate range where resonances dominate. Importantly, we can use
physical time-like scattering data to obtain ${\rm Im}\, f(s)$, which
is practical in the elastic channel thanks to Watson's final state theorem,
or when the number of open channels is not too large (typically
in the pion case one considers only $\pi\pi- K\bar K$ coupled channels~\cite{Gasser:1990bv}.)  

According to the dispersion relation~(\ref{eq:disp}), the spectral strength from 
a given range in $s$ feeds all the values of $Q^2$ (or in general, complex $t$), such that analyticity 
is guaranteed. In particular, as we will see below, resonances and $\chi$PT provide some strength also to the $1/Q^2$ tails, 
which is globally canceled by the contribution from pQCD, in harmony with the asymptotic sum rule. 
The role of the constraints following from the dispersive sum rules of
Sec.~\ref{sec:drs} on modeling of the spectral strengths has been
little considered.\footnote{To our knowledge, the only work which
  discusses the issue (for the case of the charge form factor of the
  pion) is~\cite{Donoghue:1996bt}.} Of a particular relevance are sum
rules~(\ref{eq:asr}) and (\ref{eq:thsr}), which relate to the
asymptotic behavior of form factors in the space-like region, and
provide global (i.e., involving all the values of $s$) constraints.  From this
perspective, the large mismatch of the experimental or lattice QCD
space-like form factors with the pQCD asymptotics in the available
$Q^2$ range is due to a missing (negative) strength in the corresponding
spectral density at some sufficiently high
$s$~\cite{RuizArriola:2024uub}. We will explore this issue in the
following parts of this paper.

\subsection{Asymptotic consistency \label{sec:consist}}

The leading pQCD asymptotic expressions in $Q^2$ are 
\begin{eqnarray}
&& f_p(-Q^2)\simeq c_f\frac{16\pi f_\pi^2 \alpha(-Q^2)}{Q^2} = c_f\frac{64\pi^2 f_\pi^2}{\beta_0 Q^2 \ln(Q^2/\Lambda^2)}, \label{eq:Qem}\\
&& \Theta_p(-Q^2) \simeq -4 \beta_0 \alpha(-Q^2)^2 f_\pi^2 = -\frac{64 \pi^2 f^2}{\beta_0 \ln^2 (Q^2/\Lambda^2)}, \label{eq:QTh} 
\end{eqnarray}
where the subscript $p$ indicates pQCD, $\alpha(-Q^2)=\alpha(t)=4\pi /[\beta_0 \ln(-t/\Lambda^2)]$,  
$\beta_0=\tfrac{1}{3}(11N_c-2N_f) = 9$ with 3 active flavors, and $\Lambda=225$~MeV. To treat the $F$ and $A$ channels uniformly, since here they differ by a factor of 3, we ntroduce the constant $c_f$, with $c_F=1$ and $c_A=3$. 
Analytic continuation to time-like $s$ yields, along the upper edge of the cut, 
$\alpha(s + i \epsilon)={4\pi}/({\beta_0 (L-i \pi)})$,
with the short-hand notation $L=\ln(s/\Lambda^2)$.
Then the spectral densities are
\begin{eqnarray}
&&\frac{1}{\pi} {\rm Im}\, f_p(s)=  - c_f \frac{64\pi^2 f_\pi^2}{\beta_0 s (L^2+\pi^2)}, \label{eq:imF} \\
&&\frac{1}{\pi}  {\rm Im}\, \Theta_p(s)= - \frac{128\pi^2 f_\pi^2 L}{\beta_0  (L^2+\pi^2)^2}. \label{eq:imTh}
\end{eqnarray}
The formulas are valid at $s>\Lambda_p$, a large scale where pQCD sets in. 

At first glance, there seems to be a sign clash between Eqs.~(\ref{eq:imF}) and (\ref{eq:Qem}), since the kernel $1/(s+Q^2)$ 
in the dispersion relation~(\ref{eq:disp}) is positive definite. The resolution is linked to the asymptotic sum rule, as we demonstrate below.
Changing the integration variable  
into $L=\ln(s/\Lambda^2)$, we get
\begin{eqnarray}
f_p(-Q^2)=- c_f \frac{64\pi^2 f_\pi^2}{\beta_0 Q^2} \int_{L_0}^\infty dL \frac{1}{\Lambda^2 e^L/Q^2 + 1}\frac{1}{L^2+\pi^2}, \label{eq:pQCDA}
\end{eqnarray}
where $L_0=\ln(\Lambda_p^2/\Lambda^2)$.
At large $Q^2$ the factor $1/(\Lambda^2 e^L/Q^2 + 1)$ approaches the step function $\theta[\ln(Q^2/\Lambda^2) - L]$, hence
\begin{eqnarray}
&& f_p(-Q^2) \simeq - c_f \frac{64\pi^2 f_\pi^2}{\beta_0 Q^2} \int_{L_0}^{\ln(Q^2/\Lambda^2)} \frac{dL}{L^2+\pi^2} = \nonumber \\
&& ~~~~ - c_f \frac{64\pi^2 f_\pi^2}{\beta_0 Q^2} \left [ \arctan(\ln(Q^2/\Lambda^2)/\pi)-\arctan(L_0/\pi) \right] = \nonumber \\
&& ~~~~~~~~ c_f \frac{64\pi^2 f_\pi^2}{\beta_0} \left [ \frac{1}{Q^2  \ln(Q^2/\Lambda^2)}\right ] +\frac{d_f}{Q^2} +\dots,
\end{eqnarray}
where in the last line we have expanded for asymptotic $Q^2$. The dots indicate sub-leading terms in the large $Q^2$ expansion.
Note that the first term reproduces, as a check, Eq.~(\ref{eq:Qem}) with the correct sign, 
while the second term is negative, with 
\begin{eqnarray}
d_f=c_f \frac{64\pi^2 f_\pi^2}{\beta_0}\left [ -\frac{1}{2}+ \frac{1}{\pi}\arctan(L_0/\pi) \right ] 
= -c_f \frac{64 \pi ^2 {f_\pi}^2}{{\beta_0} {L_0}} + \dots, 
\end{eqnarray}
where for clarity we have expanded for large $\Lambda_p/\Lambda$ and the dots mean terms sub-leading in $L_0$.
The $d_f/Q^2$ term (dominant over the $1/[Q^2 \ln(Q^2/\Lambda^2)]$ term) is canceled by other contributions to the asymptotics via the sum rule~(\ref{eq:asr}). 

For $\Theta$ we get with a similar calculation
\begin{eqnarray}
&& \hspace{-12mm} \Theta_p(-Q^2) \simeq - \frac{128\pi^2 f_\pi^2}{\beta_0} 
\int_{\ln(Q^2/\Lambda^2)}^\infty  \!\!\!\!\! dL \frac{L}{(L^2+\pi^2)^2} = 
- \frac{64\pi^2 f_\pi^2}{\beta_0 \ln^2(Q^2/\Lambda^2)} + \dots,
\end{eqnarray}
which agrees as a check with Eq.~(\ref{eq:QTh}). Note that in contrast to the previous $f=F, A$ case, for $\Theta$ there is no ``super-leading'' term that needs to be canceled, in accordance to the fact that 
there is no asymptotic sum rule of the form of Eq.~(\ref{eq:asr})  for $\Theta$.

\subsection{Elastic region and threshold behavior}

In the elastic region, above the two-pion and below the four-pion production threshold, $4 m_\pi^2 \le s \le 16 m_\pi^2 $, 
Watson's theorem states that
\begin{eqnarray}
f^I_{J}(s)= |f^I_{J}(s)| e^{i \delta^I_{J}(s)},
\end{eqnarray} 
where $\delta^I_{J}(s)$ are the corresponding $\pi\pi$ elastic scattering phase shifts in the isospin $I$ and spin $J$ channels. 
This  in particular implies that 
\begin{eqnarray}
  {\rm Im} f^I_{J}(s)= |f^I_{J}(s)| \sin \delta^I_{J}(s),
\end{eqnarray} 
which for attractive interactions with $ 0 \le \delta^I_{J}(s) \le
\pi$ is positive. Current analyses are consistent with the
elastic regime, to hold in practice in the extended range $4 m_\pi^2
\le s \le 4 m_K^2 $. Phenomenologically, one has resonance saturation for the three
form factors $f=F,\Theta,A$, for Breit-Wigner masses $\sqrt{s}= m_R$,
where $\delta^I_J (m_R^2) = \pi/2$ with $m_R=m_\rho, m_\sigma,
m_{f_2}$, respectively. Also, $\delta^I_J (4m_K^2) < \pi$, such that
positivity holds,
\begin{eqnarray}
  {\rm Im} A (s), {\rm Im} F (s), {\rm Im} \Theta (s) \ge 0 \,,
  \qquad 4 m_\pi^2 \le s \le 4 m_K^2  . 
\end{eqnarray}
Parametrization of the leading threshold behavior of the $\pi\!-\!\pi$ scattering amplitude
$t^I_J (s)= (e^{2i \delta^I_J (s)}-1)/\rho(s)$, where $\rho(s)=(1-4m_\pi^2/s)^\frac12$,  in 
the isospin $I$ and spin $J$ channels is
\begin{eqnarray}
t^I_{J}(s)=a^I_{J}(\tfrac{1}{4}s-m_\pi^2)^J 
\end{eqnarray} 
(which is real).
Our convention follows~\cite{Gasser:1983yg}, such that the combinations $a^I_J m^{2J}$ are dimensionless.
Using Watson's theorem, one obtains for the corresponding form factors the threshold formula to leading order,
\begin{eqnarray}
{\rm Im}\, f^I_{J}(s)=|f^I_{J}(4m_\pi^2)|\, a^I_{J}\sqrt{1-\frac{4m_\pi^2}{s}} (\tfrac{1}{4}s-m_\pi^2)^J (1 + {\cal O} (s-4m_\pi^2) ). \label{eq:anyJ}
\end{eqnarray} 

\subsection{$\chi$PT}

The NLO expressions from $\chi$PT are~\cite{Gasser:1990bv,Donoghue:1991qv}
\begin{eqnarray}
&& \frac{1}{\pi} {\rm Im}\,F_\chi(s)=\frac{1}{96\pi^2 f_\pi^2}s \left(1- \frac{4m_\pi^2}{s} \right )^{\frac{3}{2}},\\
&& \frac{1}{\pi}{\rm Im}\,\Theta_\chi(s)=\frac{1}{32\pi^2 f_\pi^2}(2m_\pi^2+s)(2s-m_\pi^2) \left(1- \frac{4m_\pi^2}{s} \right )^{\frac{1}{2}},
\label{eq:spA}
\end{eqnarray}
whereas ${\rm Im}\,A_\chi(s)$ is NNLO in chiral counting, hence
numerically tiny, and we do not include it in the forthcoming model
analysis applied to the lattice data.  The formulas are assumed to be
valid up to a scale $\Lambda_\chi$.

\section{Transverse densities from spectral strengths \label{sec:trden}}

With the help of the dispersion relation~(\ref{eq:disp}) plugged into (\ref{eq:tranden}), one can write the transverse densities
for all channels in the form~\cite{Miller:2010tz},
\begin{eqnarray}
f(b) = \frac{1}{2\pi^2} \int_{4m_\pi^2}^\infty ds K_0(b \sqrt{s}) \, {\rm Im}\, f(s), \label{eq:fbK}
\end{eqnarray}
where the order of integration over $s$ and $q_\perp$ has been
flipped, which is allowed since the integrals exist. For the cases
of $F(b)$ and $A(b)$, the positivity shown in Sec.~\ref{sec:pos} is a non-trivial condition
since, although $K_0 >0 $,  the sum rules for their spectral densities imply that
${\rm Im} F(s)$ and  ${\rm Im} A(s)$ {\it cannot} be positive. 

Vector meson dominance for the transverse $F(b)$ was exploited
in~\cite{Miller:2011du}. A recent and precise analysis considering time-like Babar data up to $s \le 9 {\rm GeV}^2$ with a modulus-phase dispersion relation~\cite{RuizArriola:2024gwb}
and consideration of sum rules~\cite{Sanchez-Puertas:2024siv}
allows a confident estimate for $b \ge 0.1{\rm fm}$~\cite{RuizArriola:2024}.

\subsection{Behavior at $b \to 0$ \label{sec:b0}}

From Eq.~(\ref{eq:fbK}) we can readily obtain the low-$b$ behavior of the 
transverse densities for the case of $f=F, A$, expanding  $K_0(b \sqrt{s} )$, where the leading term at low $b$ is $-\ln b$. 
However, this piece cancels from Eq.~(\ref{eq:fbK}) because of the 
asymptotic sum rule~(\ref{eq:asr}), as derived below. 
This generic feature is consistent with the asymptotic behavior of $f(Q^2)$ falling off faster than $1/Q^2$. 

For the transverse charge density, the singular behavior at the origin was first noticed by Gerry Miller~\cite{Miller:2009qu}, who considered a Fourier-Bessel
transform of the asymptotic $1/[Q^2 \ln(Q^2/\Lambda^2)]$ tail. Here we repeat this analysis starting from Eq.~(\ref{eq:fbK}), which
for $b\ll \sqrt{s}$ we rewrite in the form
\begin{eqnarray}
&&f(b) \simeq  c_f\frac{64 \pi^3 f_\pi^2}{\beta_0} \frac{1}{2\pi^2}\int_{\Lambda_p^2}^{1/b^2} ds \frac{1}{2}\ln (s b^2) \frac{1}{s(\ln^2(s/\Lambda)+\pi^2)} = \nonumber \\
&& ~~~~~~~ c_f \frac{49\pi f_\pi^2}{3 \beta_0} \left [ \ln \ln \left ( \frac{1}{b^2 \Lambda^2} \right ) + 
\frac{\ln(b^2 \Lambda_p^2)}{\ln (\Lambda_p^2/\Lambda^2) }\right ]. \label{eq:fbA}
\end{eqnarray}
The first term (positive singularity) is of the shape found in~\cite{Miller:2009qu}, whereas the second term (negative and dominant 
over the first term) is canceled by other contributions to 
the spectral density to satisfy the asymptotic sum rule~(\ref{eq:asr}), according to the discussion in subsection~\ref{sec:consist}.
Hence 
\begin{eqnarray}
f(b) \simeq  c_f \frac{49\pi f_\pi^2}{3 \beta_0} \ln \ln \left ( \frac{1}{b^2 \Lambda^2} \right ) . \label{eq:fasym}
\end{eqnarray}

For the scalar transverse density, obtaining the low-$b$ limit is more subtle, as the 
spectral density is not damped with the $1/s$ factor, as was the case of $f=F, A$. Scaling the integration variable by introducing
$s=S/b^2$, we can write
\begin{eqnarray}
&&\Theta(b) = -  \frac{64 \pi f_\pi^2}{\beta_0 b^2} \int_{\Lambda_p^2 b^2}^{\infty} dS K_0(\sqrt{S}) 
\frac{\ln[S/(\Lambda^2 b^2)]}{[\ln^2(S/\Lambda^2 b^2)]+\pi^2)^2} = \nonumber \\
&& ~~~~~~~ - \frac{128\pi f_\pi^2}{\beta_0}\frac{1}{b^2 \ln^3 \left ( \frac{1}{b^2 \Lambda^2}\right )} + \dots , \label{eq:fbTh}
\end{eqnarray} 
where we have expanded for $b \to \infty$ and used $\int_0^\infty dS\,K_0(\sqrt{S}) = 2$. 
The singularity is negative and integrable with $\int \! d^2 \,b$, as it should.

\subsection{Behavior at $b \to \infty$ \label{sec:binf}}

Generally, $\pi\pi$ scattering analyses find that the elastic channel practically
extends up to the $K \bar K $ threshold,  thus 
\begin{eqnarray}
  f^I_J(b) = \frac{1}{2\pi^2} \int_{4m_\pi^2}^{4m_K^2} ds  K_0(b\sqrt{s})
  |f^I_J(s)| \sin \delta^I_J (s) + {\cal O} (e^{-2 m_K b}),
\end{eqnarray}
such that for $b \gg 1/(2m_K) \sim0.2~{\rm fm}$ one has $F(b),
\Theta(b), A(b) > 0$, since the three cases correspond to attractive
interactions with $ 0 \le \delta^I_{J}(s) \le
\pi$. The very high-$b$ behavior of the transverse densities
is dictated by the $s$ behavior near the threshold $s=4m_\pi^2$.
Using Eq.~(\ref{eq:anyJ}) in Eq.~(\ref{eq:fbK}) we readily obtain the
asymptotic behavior of the transverse densities at $b\to \infty$,
\begin{eqnarray}
f^I_J(b)= m_\pi^2 |f^I_{J}(s\!=\!4m_\pi^2)|\frac{(2 J+1)!! e^{-2 b m_\pi}  [a^I_J m^{2J}]}  {2^{J+1}\pi (b m_\pi)^{J+2}}
\end{eqnarray}
For the cases of interest, and using the values of the $a^I_J$ coefficients extracted by the Bern~\cite{Colangelo:2001df} 
(upper values in the formula below)
and Madrid-Cracow~\cite{Garcia-Martin:2011iqs} (lower values) groups, we can write
\begin{eqnarray}
  && F(b)= m_\pi^2|F(s\!=\!4m_\pi^2)|\frac{ 3 e^{-2 b m_\pi} } {4\pi (b m_\pi)^3}
 \left( \begin{cases} 0.0379(5)\\ 0.0377(13)\end{cases} 
  +{\cal O}[(bm_\pi)^{-2}] \right), \nonumber \\
  && A(b)= m_\pi^2| A(s\!=\!4m_\pi^2)|\frac{5 e^{-2 b m_\pi}}
     {8\pi (b m_\pi)^4}    \left( \begin{cases} 0.00175(3)\\ 0.00178(3)\end{cases}  +{\cal O}[(bm_\pi)^{-2}] \right) , \nonumber \\
  && \Theta(b)= m_\pi^2|\Theta(s\!=\!4m_\pi^2)|\frac{e^{-2 b m_\pi}
   } {2\pi (b m_\pi)^2 }  \left( \begin{cases} 0.220(5)\\ 0.220(8)\end{cases} +{\cal O}[(bm_\pi)^{-2}] \right).
   \label{eq:b0}
\end{eqnarray}
Thus, at $b\to\infty$,
the approach of the above transverse densities to 0 is from above, which reflects the attractive nature of the $\pi\!-\!\pi$ interactions in the channels of interest, 
manifest in the positivity of $a^I_J$. Combining (\ref{eq:b0}) with Eq.~(\ref{eq:fbA},\ref{eq:fbTh}) we immediately conclude that $\Theta(b)$ must 
change sign, whereas the limits are consistent with the positive definiteness of $F(b)$ and $A(b)$. 

Using the NLO $\chi$PT formulas (\ref{eq:spA}) we get,  to the leading order in $m_\pi^2$,
\begin{eqnarray}
&& F(b)=\frac{m_\pi e^{-2 m_\pi b}}{32 \pi ^2  f_\pi^2 b^3} +{\cal O}(b^{-5}), \\
&& \Theta(b)=\frac{21 m_\pi^4 e^{-2 m_\pi b}}{32 \pi ^2 f_\pi^2 b^2}+{\cal O}(b^{-4}), \label{eq:Thbas}
\label{eq:bpA}
\end{eqnarray}
where to this order we take $|F(s\!=\!4m_\pi^2)|=1$ and $|\Theta(s\!=\!4m_\pi^2)|=6m_\pi^2$.

\section{Modeling spectral densities with resonances, pQCD, and $\chi$PT \label{sec:modspec}}

According to what has been said above, a generic model for spectral densities has the form 
\begin{eqnarray}
{\rm Im}\, f(s)=  {\rm Im}\, f_\chi(s)\theta(\Lambda_\chi^2-s) +{\rm Im}\, f_R(s)+ {\rm Im}\, f_p(s) \theta(s-\Lambda_p^2), \label{eq:mgen}
\end{eqnarray}  
with the $\chi$PT, resonance, and pQCD regions. This division is rough, as in reality there is no strict separation. For instance, in the scalar 
channel $\chi$PT merges smoothly with the $\sigma$ meson, building a continuous wide structure in $s$ from the threshold up to the $f_0(980)$ mass squared. On the other end, the towers of Regge states continue up to the perturbative region at large $s$, where they in fact mimic pQCD according to the 
parton-hadron duality principle (an example for $F$ is provided in \cite{RuizArriola:2008sq}). 
We do not enter these issues here, but just take Eq.~(\ref{eq:mgen}) as useful
to estimate the size of various contributions to the sum rules from Sec.~\ref{sec:drs}.
We take 
\begin{eqnarray}
\Lambda_\chi=0.6~{\rm GeV}, \;\;\;  \Lambda_p=3~{\rm GeV}, \label{eq:LL}
\end{eqnarray}
to evaluate the $\chi$PT and pQCD contributions to sum rules. 

For narrow resonances (which is a feature of the large-$N_c$ limit), the  resonance contribution to the spectral 
densities takes the form of sums of $\delta$ functions,
\begin{eqnarray}
\frac{1}{\pi} {\rm Im} f(s)=\sum_i a_i M_i^2 \delta(s-M_i^2), \;\;\;\;\frac{1}{\pi} {\rm Im} \Theta(s)=\sum_i b_i M_i^4 \delta(s-M_i^2),
\end{eqnarray}
where $M_i$ are the resonance masses in the appropriate spin-isospin channel and $a_i,b_i$ are their dimensionless coupling parameters.  
In approximations with just one resonance, which work phenomenologically remarkably well in the description of the data in the available $Q^2$-range, one takes the lowest mass in a given channel. In particular, for $F$, $A$, and $\Theta$ one takes $M_1=m_\rho$, $m_{f_2}$, and $m_\sigma$, respectively, with corresponding $a_1$ or $b_1$ close to unity. Even in modeling with more resonances, these lowest states are expected to be dominant. 
\begin{table}[tb]
\caption{Contributions to the sum rules. \label{tab:sr}}
\begin{center}
{\small
\begin{tabular}{l|l|l|l||c}
sum rule & $\chi$PT  & dominant res. & pQCD       & total \\ \hline
charge,  $F$, Eq.~(\ref{eq:csr}) & 0.01                    & $\sim 1$ & -0.002 & 1\\
asymp.,  $F$, Eq.~(\ref{eq:asr}) [GeV$^2$]& 0.003 & $m_\rho^2 \sim 0.6$ & -0.1 & 0\\ \hline
charge,  $A$, Eq.~(\ref{eq:csr}) & NNLO                         & $\sim 1$ & -0.005 & 1\\
asymp.,  $A$, Eq.~(\ref{eq:asr}) [GeV$^2$]& NNLO                       &  $m_{f_2}^2 \sim 1.6 $ &-0.3 & 0\\ \hline
mass,  $\Theta$, Eq.~(\ref{eq:thsr}) [GeV$^2$]& 0.03 & $m_{\sigma}^2 \sim 0.2\!-\!0.6$ & -0.02 & $2m_\pi^2= 0.02$ \\
slope,  $\Theta$, Eq.~(\ref{eq:slopeth})             & 0.1  & $ \sim 1$                                & -0.0004 & $1+{\cal O}(m_\pi^2)$
\end{tabular}
}
\end{center}
\end{table}
However, models with only one dominant resonance cannot satisfy the sum rules, as can be seen from the numbers collected in Table~\ref{tab:sr}.
The first column of the table describes the type of the sum rule. The following columns give the $\chi$PT contribution to the sum rule, the dominant resonance contribution, and the LO pQCD contribution. The last column gives the sum rule value, to which all the components should sum up.
We note that if we wish to satisfy the charge sum rules for $F$ or $A$, we need to 
take $a_1\simeq 1$, as the $\chi$PT and pQCD corrections are small, 
but then the asymptotic sum rule~(\ref{eq:asr}) is badly broken. Therefore, as argued in \cite{Broniowski:2024oyk}, one needs additional negative contributions to the spectral densities at large $s$, such that the asymptotic sum rules are mended, but the charge sum rules preserved.

\subsection{Model with two resonances \label{sec:2res}}

Here we consider a model where a second resonance, to be  treated as an effective negative strength, is included in 
each channel.\footnote{An alternative scenario based on a fractional power based on asymptotics of radial Regge trajectories~\cite{RuizArriola:2008sq} (see also \cite{RuizArriola:2024uub}) will be treated elsewhere~\cite{RuizArriola:2024}.} 
With two resonances (+pQCD +$\chi$PT), for $f=F, A$ we get from the sum rules~(\ref{eq:csr},\ref{eq:asr})
\begin{eqnarray}
a_1+a_2+c_{\rm nr}  = 1, \nonumber \\ 
{M_1^2}{a_1} + {M_2^2}{a_2} +a_{\rm nr} = 0,  \label{eq:2ch1} 
\end{eqnarray}
where nr indicates the (known) non-resonant ($\chi$PT + pQCD) contributions.
Both sum rules can obviously be satisfied now, 
with the solution
\begin{eqnarray}
&& a_1=\frac{(1-c_{\rm nr})M_2^2+a_{\rm nr}}{M_2^2-M_1^2}, \nonumber \\
&& a_2=-\frac{(1-c_{\rm nr})M_1^2+a_{\rm nr}}{M_2^2-M_1^2}, 
\end{eqnarray}
whereby
\begin{eqnarray}
f(Q^2) = \frac{(1-c_{\rm nr}) M_1^2 M_2^2 - a_{\rm nr} Q^2}{(M_1^2+Q^2)(M_2^2+Q^2)} + f_{\rm nr}(Q^2).
\end{eqnarray}
Since $f(Q^2) =1$, we have $c_{\rm nr}=f_{\rm nr}(0)$. On the other hand, in the limit of $Q^2\to\infty$
we find $a_{\rm nr}=\lim_{Q^2\to\infty}Q^2 f_{\rm nr}(Q^2)$ (cf. discussion in Sec.~\ref{sec:consist}). 
For the case of $\Theta$ we take the mass and slope sum rules,
\begin{eqnarray}
&& b_1 M_1^2+b_2 M_2^2+u_{\rm nr}  = 2m_\pi^2, \nonumber \\ 
&& {b_1} + {b_2} +v_{\rm nr} = S,  \label{eq:2ch2} 
\end{eqnarray}
where $S=1+{\cal O}(m_\pi^2/f_\pi^2)$ is the desired slope. The solution is
\begin{eqnarray}
&& b_1=- \frac{M_2^2 (S-v_{\rm nr})+u_{\rm nr}-2 m^2}{M_2^2-M_1^2}, \nonumber \\
&& b_2=-\frac{M_2^2 (S-v_{\rm nr})+u_{\rm nr}-2 m^2}{M_2^2-M_1^2}, 
\end{eqnarray}
hence
\begin{eqnarray}
&&\hspace{-4.5mm}\Theta(-Q^2)=
 \frac{(2m^2\!-\!u_{\rm nr})M_1^2 M_2^2+[(2m^2\!-\!u_{\rm nr}) (M_1^2+M_2^2)-(S\!-\!v_{\rm nr}) M_1^2 M_2^2]Q^2}{(M_1^2+Q^2)(M_2^2+Q^2)} 
\nonumber \\ 
&&~~~~~~~~~+ \Theta_{\rm nr}(-Q^2).
\end{eqnarray}
Near the origin
\begin{eqnarray}
\Theta(-Q^2)=2m^2 -u_{\rm nr} + \Theta_{\rm nr}(0) - [S-v_{\rm nr} +\Theta'_{\rm nr}(0)  ]Q^2 + {\cal O}(Q^4)
\end{eqnarray}
(the prime indicates the derivative with respect to $t$), hence $u_{\rm nr} = \Theta_{\rm nr}(0)$ and
$v_{\rm nr} =\Theta'_{\rm nr}(0)$. Asymptotically, the resonance contribution falls off as $1/Q^2$, hence the 
asymptotics of the full $\Theta$ originates from the non-resonant (pQCD) part, as discussed earlier.

We remark that the considered model, although schematic in the sense of placing all the needed negative strength in a single narrow resonance placed far away in $s$, is generic. The same features are expected from more involved modeling, where the negative strength is distributed over an extended region in $s$, or in Regge-type modeling with infinitely many resonances. A problem with more realistic modeling is the multitude of model parameters, not possible to fix with the presently available data, so much freedom/overfitting is left. 
We expect that the large negative strength in the spectral density is distributed at high $s$, beyond the range of the presently available time-like data. 

\subsection{~~$F$}

For shortage of space, we do not present the results from the model 
for the charge form factor $F$. We only mention that taking $M_1=m_\rho$ and 
$M_2$ sufficiently high, one can describe the space-like data in a satisfactory manner, in particular, the flatness of $Q^2 F(Q^2)$ reaching 
up to $Q\sim 3$~GeV. Importantly,  both $F(Q^2)$ and $F(b)$ are positive definite. The case is qualitatively the same as for $A$, 
described in detail below. 

\subsection{~~$A$}

\begin{figure}[tb]
\begin{center}
\includegraphics[width=.495\textwidth]{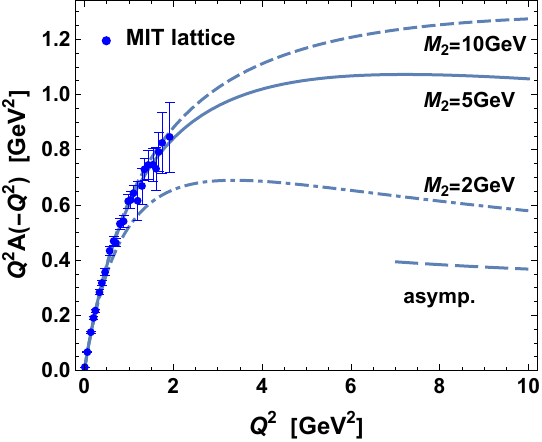} \includegraphics[width=.495\textwidth]{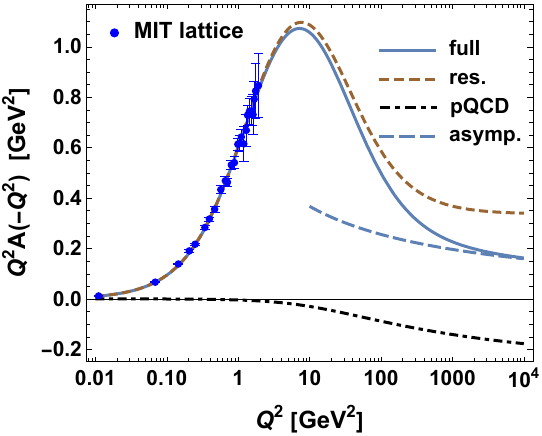}
\caption{Left: The tensor gravitational form factor of the pion multiplied by $Q^2$, 
in the model with two resonances and pQCD for $M_1=m_{f_2}=1.275$~GeV and 
several values of $M_2$ indicated in the figure.
Right: Anatomy of $Q^2 A(Q^2)$  for 
$M_1=m_{f_2}=1.275$~GeV and  $M_2=5$~GeV. Full result (solid), the resonance contribution (dashed), the pQCD contribution (dot-dashed) of Eq.~(\ref{eq:pQCDA}), and the 
asymptotic limit of Eq.~(\ref{eq:Qem}) (long dash). The lattice MIT data are from~\cite{Hackett:2023nkr}. 
\label{fig:AQ2}}
\end{center}
\end{figure}

\begin{figure}[tb]
\begin{center}
\includegraphics[width=.495\textwidth]{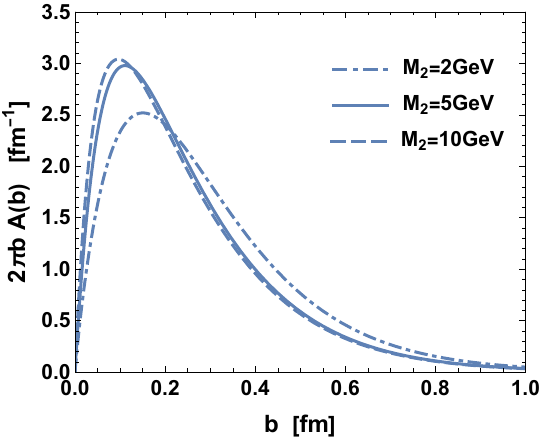} \includegraphics[width=.495\textwidth]{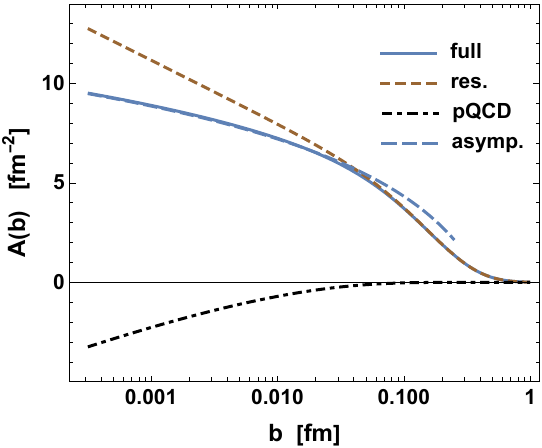}
\caption{Left: Tensor gravitational transverse density of the pion in the model with two resonances and pQCD, for $M_1=m_{f_2}=1.275$~GeV  and 
several values of $M_2$.
Right: Anatomy of $A(b)$  for 
$M_1=m_{f_2}=1.275$~GeV and  $M_2=5$~GeV. Full result (solid), the resonance contribution (dashed), the pQCD contribution (dot-dashed), and the 
asymptotic limit of Eq.~(\ref{eq:fbA}) (long dash).
\label{fig:Ab}}
\end{center}
\end{figure}

For $A$ we neglect the 
tiny NNLO $\chi$PT contribution, but retain the LO pQCD piece. The results shown in Fig.~\ref{fig:AQ2} are
for $M_1$ set to the PDG value of the $f_2(1275)$ meson and for several large values of $M_2$.
The comparison with the MIT lattice QCD data~\cite{Hackett:2023nkr} in the left panel shows that values of $M_2$ from $\sim 5$~GeV upwards are 
admissible, whereas $M_2=2$~GeV is visibly too low. 
The values of the couplings for $M_2= 5$~GeV are $a_1=1.06$ and $a_2=-0.056$, while 
$c_{\rm nr}=-0.004$, in satisfaction of the charge sum rule. Of course, the coupling of the second resonance is negative, and its contribution 
to the asymptotic sum rule is large due to the large value of $M_2 > M_1$.
The anatomy of $A$ is shown in the right panel of Fig.~\ref{fig:AQ2}. We note a very slow approach to the asymptotic pQCD limit, which reflects the large value of $M_2$ required by the data. Asymptotically, the $1/Q^2$ tail contribution from the resonances is exactly canceled by 
the pQCD contribution of Eq.~(\ref{eq:pQCDA}) via sum rule~(\ref{eq:asr}), leaving the {\em proper} pQCD asymptotic tail of 
Eq.~(\ref{eq:Qem}).
In Fig.~\ref{fig:Ab} we present the corresponding transverse density (multiplied with $2\pi b$). We note that it satisfies the positivity 
condition $A(b)>0$ (cf. Sec.~\ref{sec:pos}). The left panel shows the comparison of several values of $M_2$, whereas the right panel shows 
the various components of $A(b)$, with the cancelation at low $b$ as discussed around Eq.~(\ref{eq:fbA}).

\subsection{~~$\Theta$}

\begin{figure}[tb]
\begin{center}
\includegraphics[width=.495\textwidth]{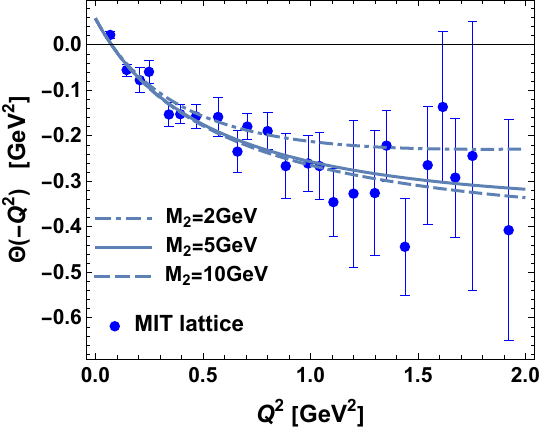} \includegraphics[width=.495\textwidth]{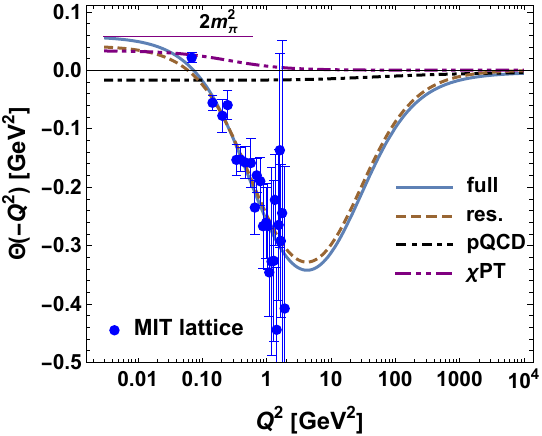}
\caption{Left: The scalar gravitational form factor of the pion 
in the model with two resonances, pQCD, and $\chi$PT for $M_1=m_{\sigma}=800$~MeV and 
several values of $M_2$.  
Right: Anatomy of $\Theta(Q^2)$  for  $M_1=m_{\sigma}=800$~MeV and  $M_2=5$~GeV. Full result (solid), the resonance contribution (dashed), the pQCD contribution (dot-dashed) of Eq.~(\ref{eq:pQCDA}), the $\chi$PT contribution (dot-dot-dashed), and the 
asymptotic limit of Eq.~(\ref{eq:Qem}) (long dash). 
The lattice MIT data are from~\cite{Hackett:2023nkr}. \label{fig:Th_lin}}
\end{center}
\end{figure}

 \begin{figure}[tb]
\begin{center}
\includegraphics[width=.495\textwidth]{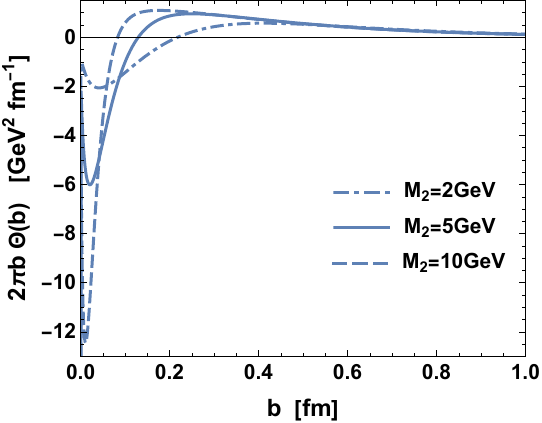} \includegraphics[width=.495\textwidth]{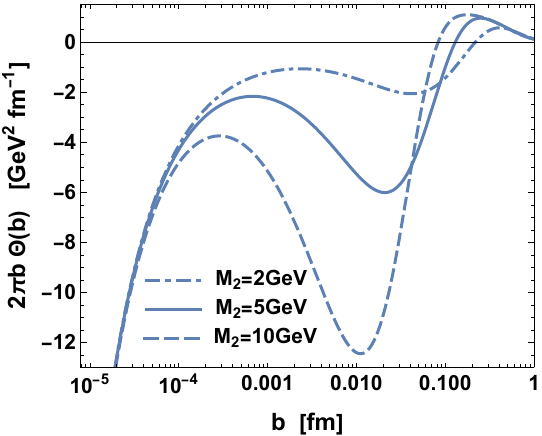}
\caption{Left: Scalar gravitational transverse density of the pion in the model with two resonances and pQCD, 
for $M_1=m_{\sigma}=800$~MeV  and 
several values of $M_2$.
Right: Same as the right panel but for $b$ in the logarithmic scale.
\label{fig:Thb}}
\end{center}
\end{figure}

\begin{figure}[tb]
\begin{center}
\includegraphics[width=.48\textwidth]{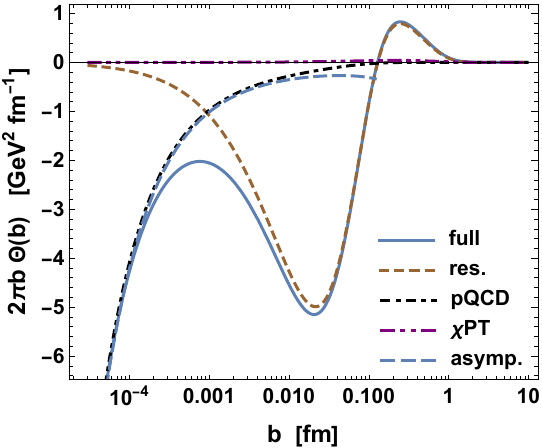} \includegraphics[width=.51\textwidth]{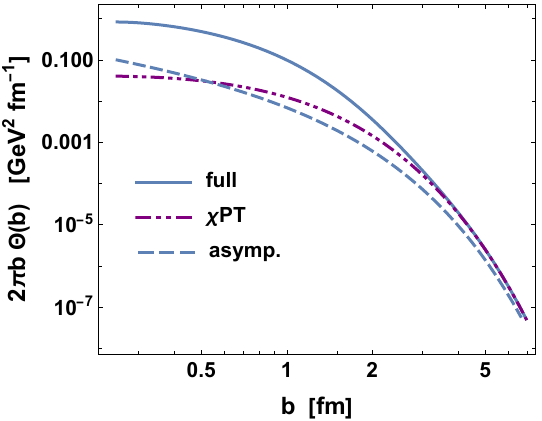}
\caption{Left: Anatomy of $2\pi b\,\Theta(b)$
for $M_1=m_{\sigma}=800$~MeV  and $M_2=5$~GeV. The asymptotics at low $b$ is from Eq.~(\ref{eq:fbTh}).
Right: Same as the right panel but with the focus on large $b$ and asymptotics from Eq.~(\ref{eq:Thbas}).
\label{fig:Thban}}
\end{center}
\end{figure}

The results for the scalar gravitational form factor in the model with two resonances are 
displayed  in Fig.~\ref{fig:Th_lin}. We use an effective sigma meson of mass $m_\sigma=800$~MeV and several values of 
$M_2$. The slope is set to $S=0.9$~\cite{Broniowski:2024oyk,RuizArriola:2024uub}.  As is apparent from the plot
in the left panel, the lattice data are yet not accurate enough to 
discriminate between different values of $M_2$, with values from $\sim2$~GeV upwards admissible. 
The right panel shows the anatomy of $\Theta(-Q^2)$.
In Fig.~\ref{fig:Thb} we show $2\pi b \, \Theta(b)$. We note a singularity at $b\to 0$, the crossing of zero at $b\sim 0.2$~fm, and a 
non-monotonic behavior a lower values of $b$. The anatomy is displayed in Fig.~\ref{fig:Thban}. We note that in the 
range $\sim 0.01-1$~fm the resonance contribution dominates.

\subsection{Transverse pressure}

Finally, we look at one of the mechanistic properties of the pion, namely, the transverse pressure, obtained from the two 
gravitational transverse densities. 
From Eq.~(\ref{eq:press}), by contracting with $\delta_{ij}$ ($i,j=1,2$), one readily finds
\begin{eqnarray}
&& 2 P^+ p(b)=-\frac{1}{4} \int \frac{d^2q_\perp}{(2\pi)^2} e^{-i q_\perp \cdot b}  q_\perp^2 D(-q_\perp^2) = \nonumber \\
&& ~~~~ -\frac{1}{6} \int \frac{d^2q_\perp}{(2\pi)^2} e^{-i q_\perp \cdot b}  
 \left [ \Theta(-q_\perp^2) - ( 2 m_\pi^2 +\tfrac{1}{2}\, q_\perp^2) A(-q_\perp^2) \right], \label{eq:ppp}
\end{eqnarray}  
where Eq.~(\ref{eq:Drel}) has been used. With the behavior of $\Theta(-q_\perp^2)$ and $A(-q_\perp^2)$ near 0 we find that 
$\int d^2b\, p(b)=0$. Also, $2 P^+ \int d^2b\, p(b)=\tfrac{1}{4} D(0)$, which is the transverse version of the relation given in~\cite{Polyakov:2018zvc}.
From Eq.~(\ref{eq:ppp}) and the $b$-representations of the form factors we find 
\begin{eqnarray}
&& 2 P^+ p(b)=-\frac{1}{6} \Theta(b) +\frac{m_\pi^2}{3} A(b) - \frac{1}{12}A_1(b),
\end{eqnarray}  
where
\begin{eqnarray}
&& A_1(b) = - \int \frac{d^2q_\perp}{(2\pi)^2} e^{-i q_\perp \cdot b} q_\perp^2 A(-q_\perp^2) =
\int \frac{d^2q_\perp}{(2\pi)^2} e^{-i q_\perp \cdot b} \times \label{eq:fbK3} \\
&& ~~~\frac{1}{\pi} \int_{4m_\pi^2}^\infty ds \left [ \frac{s}{s+q_\perp^2} -1 \right ] {\rm Im} \, A(s)=
\frac{1}{2\pi^2} \int_{4m_\pi^2}^\infty ds \,K_0(b \sqrt{s}) \,s \, {\rm Im}\,A(s), \nonumber
\end{eqnarray}
where the term with -1 in the square bracket (that would yield a singular
contribution proportional to $\delta^2(b)$) cancels thanks to the asymptotic sum 
rule~(\ref{eq:asr}).
Derivation as for Eq.~(\ref{eq:fbTh}) gives at low $b$ the singularity
\begin{eqnarray}
A_1(b)  = - \frac{196\pi f_\pi^2}{\beta_0}\frac{1}{b^2 \ln^2 \left ( \frac{1}{b^2 \Lambda^2}\right )} + \dots , \label{eq:fbA1}
\end{eqnarray} 
which is one power of the log stronger 
than the singularity in $\Theta(b)$ of Eq.~(\ref{eq:fbK}). From here we can see that $p(b)$ tends to positive infinity 
in the $b\to 0$ limit,
\begin{eqnarray}
2\pi p(b)=\frac{49\pi f_\pi^2}{3\beta_0}\frac{1}{b^2 \ln^2 \left ( \frac{1}{b^2 \Lambda^2}\right )} +{\cal O}[1/(b^2  \ln^3 b)] 
\end{eqnarray}
 On the other end, at $b\to\infty$ it approaches 0 from below, hence must change sign. This is in compliance with stability, where a positive pressure in the inner region is balanced with a negative pressure outside.

\begin{figure}[tb]
\begin{center}
\includegraphics[width=.51\textwidth]{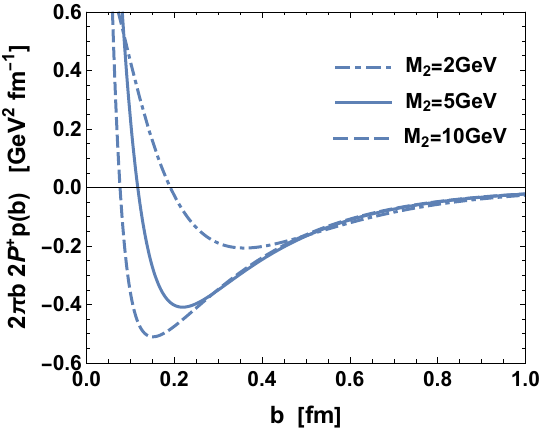} \includegraphics[width=.48\textwidth]{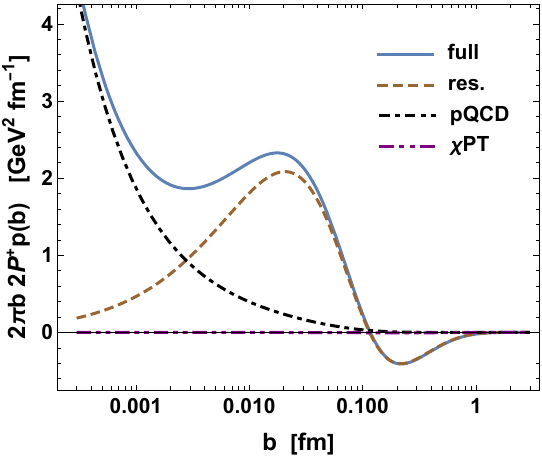}
\caption{Left: Transverse pressure in the pion, $2\pi b\,2P^+ p(b)$, in the model 
with two resonances per  $A$ and  $\Theta$, pQCD, and $\chi$PT. 
We take $m_{f_2}=1.275$~GeV, $m_{\sigma}=800$~MeV, and several values of $M_2$, same for both channels.
Right: Anatomy of the transverse pressure for $M_2=5$~GeV.
\label{fig:press}}
\end{center}
\end{figure}

The above statements concerning pressure arrival are general. We now pass to an illustration 
in the model used in the previous sections, with two resonances in each channel. 
The results are shown in Fig.~\ref{fig:press}. We note the dominance of the resonance contribution in the 
range  $\sim 0.01-1$~fm, and a remarkable smallness  of the $\chi$PT component, which nevertheless becomes
dominant at large $b$ (above $\sim 3$~fm), as required by the limits~(\ref{eq:b0}).

\section{Conclusions \label{sec:cls}}

\begin{table}[t]
\caption{Summary of the gravitational properties of the pion. \label{tab:sum}}
\begin{center}
{\small
\begin{tabular}{llll}
quantity &  low limit & intermediate range   & high limit \\ \hline
${\rm Im}\, F(s), {\rm Im}\, A(s)$  & $+$ Eq.~(\ref{eq:anyJ}) & changes sign   & $-$ Eq.~(\ref{eq:imF} \\
${\rm Im}\, \Theta(s)$  & $+$ Eq.~(\ref{eq:anyJ}) & changes sign   & $-$ Eq.~(\ref{eq:imTh} \\
$F(-Q^2),  A(-Q^2)$  & $+$ $1$ &   & $+$ Eq.~(\ref{eq:Qem} \\
$\Theta(-Q^2)$  & $+$ $2m_\pi^2$ & changes sign  & $-$ Eq.~(\ref{eq:QTh} \\
$F(b),  A(b)$  & $+$ Eq.~(\ref{eq:fasym}) &  positive definite & $+$ Eq.~(\ref{eq:b0} \\
$\Theta(b)$  & $-$ Eq.~(\ref{eq:fbTh}) & changes sign  & $+$ Eq.~(\ref{eq:b0} 
\end{tabular}
}
\end{center}
\end{table}

In this contribution we have reviewed some general features of the
gravitational form factors and the related transverse densities of the
pion. We have used analyticity and the available information from $\pi-\pi$ scattering
and pQCD to draw general conclusions on the behavior of these quantities. 
In particular, the scalar gravitational transverse density (related to the trace anomaly) must 
change sign as a function of the transverse coordinate $b$. On the other hand, the tensor 
gravitational transverse density (similarly to the electromagnetic charge case) is positive definite for all values of $b$, 
as deduced in the light-front quantization framework in the $A^+=0$ gauge. This positivity feature allows for a probabilistic interpretation. 

The basic properties of the spectral densities, form factors for space-like momenta, and the transverse densities are collected 
in Table~\ref{tab:sum}, with references to the explicit formulas in the text. The signs of the low- and high values of the arguments are indicated. 
For the spectral densities, the ``low" limit means the behavior right from the $2\pi$ production threshold, while ``high" means the asymptotic limit. 
For the other cases ``low" means at zero.

We have also discussed the implications of the recent MIT lattice QCD analysis of the
pion GFFs in the space-like region for $0< Q^2 < 2~{\rm GeV}^2$, which roughly maps into
the $b > 0.1~{\rm fm}$ region, and show that the data can be well described within the meson dominance approach. 
While a single resonance per channel suffices, 
to satisfy the sum rules following from the short-distance constraints of pQCD (the large-$Q^2$ behavior), we have added 
the needed negative contribution to the spectral densities in the form of a delta function and argued it has to appear at sufficiently high $s$, 
at least a few GeV${}^2$, not to spoil the agreement with the lattice data.

\newpage

We are grateful to the authors of Ref.~\cite{Hackett:2023nkr} for providing us the data used in the figures. 
ERA was supported by
Spanish MINECO and European FEDER funds grant and Project No.
PID2023-147072NB-I00 funded by MCIN/AEI/10.13039/501100011033,
and by the Junta de Andaluc\'\i a grant FQM-225.

\bibliography{newrefs,refs-barpi,Ref}

\begin{thebibliography}{62}%
\makeatletter
\providecommand \@ifxundefined [1]{%
 \@ifx{#1\undefined}
}%
\providecommand \@ifnum [1]{%
 \ifnum #1\expandafter \@firstoftwo
 \else \expandafter \@secondoftwo
 \fi
}%
\providecommand \@ifx [1]{%
 \ifx #1\expandafter \@firstoftwo
 \else \expandafter \@secondoftwo
 \fi
}%
\providecommand \natexlab [1]{#1}%
\providecommand \enquote  [1]{``#1''}%
\providecommand \bibnamefont  [1]{#1}%
\providecommand \bibfnamefont [1]{#1}%
\providecommand \citenamefont [1]{#1}%
\providecommand \href@noop [0]{\@secondoftwo}%
\providecommand \href [0]{\begingroup \@sanitize@url \@href}%
\providecommand \@href[1]{\@@startlink{#1}\@@href}%
\providecommand \@@href[1]{\endgroup#1\@@endlink}%
\providecommand \@sanitize@url [0]{\catcode `\\12\catcode `\$12\catcode
  `\&12\catcode `\#12\catcode `\^12\catcode `\_12\catcode `\%12\relax}%
\providecommand \@@startlink[1]{}%
\providecommand \@@endlink[0]{}%
\providecommand \url  [0]{\begingroup\@sanitize@url \@url }%
\providecommand \@url [1]{\endgroup\@href {#1}{\urlprefix }}%
\providecommand \urlprefix  [0]{URL }%
\providecommand \Eprint [0]{\href }%
\providecommand \doibase [0]{http://dx.doi.org/}%
\providecommand \selectlanguage [0]{\@gobble}%
\providecommand \bibinfo  [0]{\@secondoftwo}%
\providecommand \bibfield  [0]{\@secondoftwo}%
\providecommand \translation [1]{[#1]}%
\providecommand \BibitemOpen [0]{}%
\providecommand \bibitemStop [0]{}%
\providecommand \bibitemNoStop [0]{.\EOS\space}%
\providecommand \EOS [0]{\spacefactor3000\relax}%
\providecommand \BibitemShut  [1]{\csname bibitem#1\endcsname}%
\let\auto@bib@innerbib\@empty
\bibitem [{\citenamefont {Polyakov}\ and\ \citenamefont
  {Weiss}(1999)}]{Polyakov:1999gs}%
  \BibitemOpen
  \bibfield  {author} {\bibinfo {author} {\bibfnamefont {M.~V.}\ \bibnamefont
  {Polyakov}}\ and\ \bibinfo {author} {\bibfnamefont {C.}~\bibnamefont
  {Weiss}},\ }\href {\doibase 10.1103/PhysRevD.60.114017} {\bibfield  {journal}
  {\bibinfo  {journal} {Phys. Rev. D}\ }\textbf {\bibinfo {volume} {60}},\
  \bibinfo {pages} {114017} (\bibinfo {year} {1999})},\ \Eprint
  {http://arxiv.org/abs/hep-ph/9902451} {arXiv:hep-ph/9902451} \BibitemShut
  {NoStop}%
\bibitem [{\citenamefont {Polyakov}(2003)}]{Polyakov:2002yz}%
  \BibitemOpen
  \bibfield  {author} {\bibinfo {author} {\bibfnamefont {M.~V.}\ \bibnamefont
  {Polyakov}},\ }\href {\doibase 10.1016/S0370-2693(03)00036-4} {\bibfield
  {journal} {\bibinfo  {journal} {Phys. Lett. B}\ }\textbf {\bibinfo {volume}
  {555}},\ \bibinfo {pages} {57} (\bibinfo {year} {2003})},\ \Eprint
  {http://arxiv.org/abs/hep-ph/0210165} {arXiv:hep-ph/0210165} \BibitemShut
  {NoStop}%
\bibitem [{\citenamefont {Polyakov}\ and\ \citenamefont
  {Schweitzer}(2018)}]{Polyakov:2018zvc}%
  \BibitemOpen
  \bibfield  {author} {\bibinfo {author} {\bibfnamefont {M.~V.}\ \bibnamefont
  {Polyakov}}\ and\ \bibinfo {author} {\bibfnamefont {P.}~\bibnamefont
  {Schweitzer}},\ }\href {\doibase 10.1142/S0217751X18300259} {\bibfield
  {journal} {\bibinfo  {journal} {Int. J. Mod. Phys. A}\ }\textbf {\bibinfo
  {volume} {33}},\ \bibinfo {pages} {1830025} (\bibinfo {year} {2018})},\
  \Eprint {http://arxiv.org/abs/1805.06596} {arXiv:1805.06596 [hep-ph]}
  \BibitemShut {NoStop}%
\bibitem [{\citenamefont {Carruthers}(1971)}]{Carruthers:1971vz}%
  \BibitemOpen
  \bibfield  {author} {\bibinfo {author} {\bibfnamefont {P.}~\bibnamefont
  {Carruthers}},\ }\href {\doibase 10.1016/0370-1573(71)90010-X} {\bibfield
  {journal} {\bibinfo  {journal} {Phys. Rept.}\ }\textbf {\bibinfo {volume}
  {1}},\ \bibinfo {pages} {1} (\bibinfo {year} {1971})}\BibitemShut {NoStop}%
\bibitem [{\citenamefont {Sharp}\ and\ \citenamefont
  {Wagner}(1963)}]{sharp1963asymptotic}%
  \BibitemOpen
  \bibfield  {author} {\bibinfo {author} {\bibfnamefont {D.~H.}\ \bibnamefont
  {Sharp}}\ and\ \bibinfo {author} {\bibfnamefont {W.~G.}\ \bibnamefont
  {Wagner}},\ }\href@noop {} {\bibfield  {journal} {\bibinfo  {journal}
  {Physical Review}\ }\textbf {\bibinfo {volume} {131}},\ \bibinfo {pages}
  {2226} (\bibinfo {year} {1963})}\BibitemShut {NoStop}%
\bibitem [{\citenamefont {Pagels}(1966)}]{Pagels:1966zza}%
  \BibitemOpen
  \bibfield  {author} {\bibinfo {author} {\bibfnamefont {H.}~\bibnamefont
  {Pagels}},\ }\href {\doibase 10.1103/PhysRev.144.1250} {\bibfield  {journal}
  {\bibinfo  {journal} {Phys. Rev.}\ }\textbf {\bibinfo {volume} {144}},\
  \bibinfo {pages} {1250} (\bibinfo {year} {1966})}\BibitemShut {NoStop}%
\bibitem [{\citenamefont {Raman}(1971)}]{Raman:1971jg}%
  \BibitemOpen
  \bibfield  {author} {\bibinfo {author} {\bibfnamefont {K.}~\bibnamefont
  {Raman}},\ }\href {\doibase 10.1103/PhysRevD.4.476} {\bibfield  {journal}
  {\bibinfo  {journal} {Phys. Rev. D}\ }\textbf {\bibinfo {volume} {4}},\
  \bibinfo {pages} {476} (\bibinfo {year} {1971})}\BibitemShut {NoStop}%
\bibitem [{\citenamefont {Truong}\ and\ \citenamefont
  {Willey}(1989)}]{Truong:1989my}%
  \BibitemOpen
  \bibfield  {author} {\bibinfo {author} {\bibfnamefont {T.~N.}\ \bibnamefont
  {Truong}}\ and\ \bibinfo {author} {\bibfnamefont {R.~S.}\ \bibnamefont
  {Willey}},\ }\href {\doibase 10.1103/PhysRevD.40.3635} {\bibfield  {journal}
  {\bibinfo  {journal} {Phys. Rev. D}\ }\textbf {\bibinfo {volume} {40}},\
  \bibinfo {pages} {3635} (\bibinfo {year} {1989})}\BibitemShut {NoStop}%
\bibitem [{\citenamefont {Gasser}\ and\ \citenamefont
  {Meissner}(1991)}]{Gasser:1990bv}%
  \BibitemOpen
  \bibfield  {author} {\bibinfo {author} {\bibfnamefont {J.}~\bibnamefont
  {Gasser}}\ and\ \bibinfo {author} {\bibfnamefont {U.~G.}\ \bibnamefont
  {Meissner}},\ }\href {\doibase 10.1016/0550-3213(91)90460-F} {\bibfield
  {journal} {\bibinfo  {journal} {Nucl. Phys. B}\ }\textbf {\bibinfo {volume}
  {357}},\ \bibinfo {pages} {90} (\bibinfo {year} {1991})}\BibitemShut
  {NoStop}%
\bibitem [{\citenamefont {Donoghue}\ and\ \citenamefont
  {Leutwyler}(1991)}]{Donoghue:1991qv}%
  \BibitemOpen
  \bibfield  {author} {\bibinfo {author} {\bibfnamefont {J.~F.}\ \bibnamefont
  {Donoghue}}\ and\ \bibinfo {author} {\bibfnamefont {H.}~\bibnamefont
  {Leutwyler}},\ }\href {\doibase 10.1007/BF01560453} {\bibfield  {journal}
  {\bibinfo  {journal} {Z. Phys. C}\ }\textbf {\bibinfo {volume} {52}},\
  \bibinfo {pages} {343} (\bibinfo {year} {1991})}\BibitemShut {NoStop}%
\bibitem [{\citenamefont {Hackett}\ \emph {et~al.}(2023)\citenamefont
  {Hackett}, \citenamefont {Oare}, \citenamefont {Pefkou},\ and\ \citenamefont
  {Shanahan}}]{Hackett:2023nkr}%
  \BibitemOpen
  \bibfield  {author} {\bibinfo {author} {\bibfnamefont {D.~C.}\ \bibnamefont
  {Hackett}}, \bibinfo {author} {\bibfnamefont {P.~R.}\ \bibnamefont {Oare}},
  \bibinfo {author} {\bibfnamefont {D.~A.}\ \bibnamefont {Pefkou}}, \ and\
  \bibinfo {author} {\bibfnamefont {P.~E.}\ \bibnamefont {Shanahan}},\ }\href
  {\doibase 10.1103/PhysRevD.108.114504} {\bibfield  {journal} {\bibinfo
  {journal} {Phys. Rev. D}\ }\textbf {\bibinfo {volume} {108}},\ \bibinfo
  {pages} {114504} (\bibinfo {year} {2023})},\ \Eprint
  {http://arxiv.org/abs/2307.11707} {arXiv:2307.11707 [hep-lat]} \BibitemShut
  {NoStop}%
\bibitem [{\citenamefont {Pefkou}(2023)}]{Pefkou:2023okb}%
  \BibitemOpen
  \bibfield  {author} {\bibinfo {author} {\bibfnamefont {D.~A.}\ \bibnamefont
  {Pefkou}},\ }\emph {\bibinfo {title} {{Gravitational form factors of hadrons
  from lattice QCD}}},\ \href@noop {} {Ph.D. thesis},\ \bibinfo  {school} {MIT}
  (\bibinfo {year} {2023})\BibitemShut {NoStop}%
\bibitem [{\citenamefont {Brommel}(2007)}]{Brommel:2007zz}%
  \BibitemOpen
  \bibfield  {author} {\bibinfo {author} {\bibfnamefont {D.}~\bibnamefont
  {Brommel}},\ }\emph {\bibinfo {title} {{Pion Structure from the Lattice}}},\
  \href {\doibase 10.3204/DESY-THESIS-2007-023} {Ph.D. thesis},\ \bibinfo
  {school} {Regensburg U.} (\bibinfo {year} {2007})\BibitemShut {NoStop}%
\bibitem [{\citenamefont {Br\"ommel}\ \emph {et~al.}(2008)\citenamefont
  {Br\"ommel} \emph {et~al.}}]{QCDSF:2007ifr}%
  \BibitemOpen
  \bibfield  {author} {\bibinfo {author} {\bibfnamefont {D.}~\bibnamefont
  {Br\"ommel}} \emph {et~al.} (\bibinfo {collaboration} {QCDSF, UKQCD}),\
  }\href {\doibase 10.1103/PhysRevLett.101.122001} {\bibfield  {journal}
  {\bibinfo  {journal} {Phys. Rev. Lett.}\ }\textbf {\bibinfo {volume} {101}},\
  \bibinfo {pages} {122001} (\bibinfo {year} {2008})},\ \Eprint
  {http://arxiv.org/abs/0708.2249} {arXiv:0708.2249 [hep-lat]} \BibitemShut
  {NoStop}%
\bibitem [{\citenamefont {Delmar}\ \emph {et~al.}(2024)\citenamefont {Delmar},
  \citenamefont {Alexandrou}, \citenamefont {Bacchio}, \citenamefont {Clo\"et},
  \citenamefont {Constantinou},\ and\ \citenamefont
  {Koutsou}}]{Delmar:2024vxn}%
  \BibitemOpen
  \bibfield  {author} {\bibinfo {author} {\bibfnamefont {J.}~\bibnamefont
  {Delmar}}, \bibinfo {author} {\bibfnamefont {C.}~\bibnamefont {Alexandrou}},
  \bibinfo {author} {\bibfnamefont {S.}~\bibnamefont {Bacchio}}, \bibinfo
  {author} {\bibfnamefont {I.}~\bibnamefont {Clo\"et}}, \bibinfo {author}
  {\bibfnamefont {M.}~\bibnamefont {Constantinou}}, \ and\ \bibinfo {author}
  {\bibfnamefont {G.}~\bibnamefont {Koutsou}},\ }in\ \href@noop {} {\emph
  {\bibinfo {booktitle} {{40th International Symposium on Lattice Field
  Theory}}}}\ (\bibinfo {year} {2024})\ \Eprint
  {http://arxiv.org/abs/2401.04080} {arXiv:2401.04080 [hep-lat]} \BibitemShut
  {NoStop}%
\bibitem [{\citenamefont {Shanahan}\ and\ \citenamefont
  {Detmold}(2019)}]{Shanahan:2018pib}%
  \BibitemOpen
  \bibfield  {author} {\bibinfo {author} {\bibfnamefont {P.~E.}\ \bibnamefont
  {Shanahan}}\ and\ \bibinfo {author} {\bibfnamefont {W.}~\bibnamefont
  {Detmold}},\ }\href {\doibase 10.1103/PhysRevD.99.014511} {\bibfield
  {journal} {\bibinfo  {journal} {Phys. Rev. D}\ }\textbf {\bibinfo {volume}
  {99}},\ \bibinfo {pages} {014511} (\bibinfo {year} {2019})},\ \Eprint
  {http://arxiv.org/abs/1810.04626} {arXiv:1810.04626 [hep-lat]} \BibitemShut
  {NoStop}%
\bibitem [{\citenamefont {Wang}\ \emph
  {et~al.}(2024{\natexlab{a}})\citenamefont {Wang}, \citenamefont {He},
  \citenamefont {Wang}, \citenamefont {Draper}, \citenamefont {Liang},
  \citenamefont {Liu},\ and\ \citenamefont {Yang}}]{Wang:2024lrm}%
  \BibitemOpen
  \bibfield  {author} {\bibinfo {author} {\bibfnamefont {B.}~\bibnamefont
  {Wang}}, \bibinfo {author} {\bibfnamefont {F.}~\bibnamefont {He}}, \bibinfo
  {author} {\bibfnamefont {G.}~\bibnamefont {Wang}}, \bibinfo {author}
  {\bibfnamefont {T.}~\bibnamefont {Draper}}, \bibinfo {author} {\bibfnamefont
  {J.}~\bibnamefont {Liang}}, \bibinfo {author} {\bibfnamefont {K.-F.}\
  \bibnamefont {Liu}}, \ and\ \bibinfo {author} {\bibfnamefont {Y.-B.}\
  \bibnamefont {Yang}} (\bibinfo {collaboration} {\ensuremath{\chi}QCD}),\
  }\href {\doibase 10.1103/PhysRevD.109.094504} {\bibfield  {journal} {\bibinfo
   {journal} {Phys. Rev. D}\ }\textbf {\bibinfo {volume} {109}},\ \bibinfo
  {pages} {094504} (\bibinfo {year} {2024}{\natexlab{a}})},\ \Eprint
  {http://arxiv.org/abs/2401.05496} {arXiv:2401.05496 [hep-lat]} \BibitemShut
  {NoStop}%
\bibitem [{\citenamefont {Masuda}\ \emph {et~al.}(2016)\citenamefont {Masuda}
  \emph {et~al.}}]{Belle:2015oin}%
  \BibitemOpen
  \bibfield  {author} {\bibinfo {author} {\bibfnamefont {M.}~\bibnamefont
  {Masuda}} \emph {et~al.} (\bibinfo {collaboration} {Belle}),\ }\href
  {\doibase 10.1103/PhysRevD.93.032003} {\bibfield  {journal} {\bibinfo
  {journal} {Phys. Rev. D}\ }\textbf {\bibinfo {volume} {93}},\ \bibinfo
  {pages} {032003} (\bibinfo {year} {2016})},\ \Eprint
  {http://arxiv.org/abs/1508.06757} {arXiv:1508.06757 [hep-ex]} \BibitemShut
  {NoStop}%
\bibitem [{\citenamefont {Kumano}\ \emph {et~al.}(2018)\citenamefont {Kumano},
  \citenamefont {Song},\ and\ \citenamefont {Teryaev}}]{Kumano:2017lhr}%
  \BibitemOpen
  \bibfield  {author} {\bibinfo {author} {\bibfnamefont {S.}~\bibnamefont
  {Kumano}}, \bibinfo {author} {\bibfnamefont {Q.-T.}\ \bibnamefont {Song}}, \
  and\ \bibinfo {author} {\bibfnamefont {O.~V.}\ \bibnamefont {Teryaev}},\
  }\href {\doibase 10.1103/PhysRevD.97.014020} {\bibfield  {journal} {\bibinfo
  {journal} {Phys. Rev. D}\ }\textbf {\bibinfo {volume} {97}},\ \bibinfo
  {pages} {014020} (\bibinfo {year} {2018})},\ \Eprint
  {http://arxiv.org/abs/1711.08088} {arXiv:1711.08088 [hep-ph]} \BibitemShut
  {NoStop}%
\bibitem [{\citenamefont {Broniowski}\ \emph {et~al.}(2008)\citenamefont
  {Broniowski}, \citenamefont {Ruiz~Arriola},\ and\ \citenamefont
  {Golec-Biernat}}]{Broniowski:2007si}%
  \BibitemOpen
  \bibfield  {author} {\bibinfo {author} {\bibfnamefont {W.}~\bibnamefont
  {Broniowski}}, \bibinfo {author} {\bibfnamefont {E.}~\bibnamefont
  {Ruiz~Arriola}}, \ and\ \bibinfo {author} {\bibfnamefont {K.}~\bibnamefont
  {Golec-Biernat}},\ }\href {\doibase 10.1103/PhysRevD.77.034023} {\bibfield
  {journal} {\bibinfo  {journal} {Phys. Rev. D}\ }\textbf {\bibinfo {volume}
  {77}},\ \bibinfo {pages} {034023} (\bibinfo {year} {2008})},\ \Eprint
  {http://arxiv.org/abs/0712.1012} {arXiv:0712.1012 [hep-ph]} \BibitemShut
  {NoStop}%
\bibitem [{\citenamefont {Broniowski}\ and\ \citenamefont
  {Ruiz~Arriola}(2008)}]{Broniowski:2008hx}%
  \BibitemOpen
  \bibfield  {author} {\bibinfo {author} {\bibfnamefont {W.}~\bibnamefont
  {Broniowski}}\ and\ \bibinfo {author} {\bibfnamefont {E.}~\bibnamefont
  {Ruiz~Arriola}},\ }\href {\doibase 10.1103/PhysRevD.78.094011} {\bibfield
  {journal} {\bibinfo  {journal} {Phys. Rev. D}\ }\textbf {\bibinfo {volume}
  {78}},\ \bibinfo {pages} {094011} (\bibinfo {year} {2008})},\ \Eprint
  {http://arxiv.org/abs/0809.1744} {arXiv:0809.1744 [hep-ph]} \BibitemShut
  {NoStop}%
\bibitem [{\citenamefont {Frederico}\ \emph {et~al.}(2009)\citenamefont
  {Frederico}, \citenamefont {Pace}, \citenamefont {Pasquini},\ and\
  \citenamefont {Salme}}]{Frederico:2009fk}%
  \BibitemOpen
  \bibfield  {author} {\bibinfo {author} {\bibfnamefont {T.}~\bibnamefont
  {Frederico}}, \bibinfo {author} {\bibfnamefont {E.}~\bibnamefont {Pace}},
  \bibinfo {author} {\bibfnamefont {B.}~\bibnamefont {Pasquini}}, \ and\
  \bibinfo {author} {\bibfnamefont {G.}~\bibnamefont {Salme}},\ }\href
  {\doibase 10.1103/PhysRevD.80.054021} {\bibfield  {journal} {\bibinfo
  {journal} {Phys. Rev. D}\ }\textbf {\bibinfo {volume} {80}},\ \bibinfo
  {pages} {054021} (\bibinfo {year} {2009})},\ \Eprint
  {http://arxiv.org/abs/0907.5566} {arXiv:0907.5566 [hep-ph]} \BibitemShut
  {NoStop}%
\bibitem [{\citenamefont {Masjuan}\ \emph {et~al.}(2013)\citenamefont
  {Masjuan}, \citenamefont {Ruiz~Arriola},\ and\ \citenamefont
  {Broniowski}}]{Masjuan:2012sk}%
  \BibitemOpen
  \bibfield  {author} {\bibinfo {author} {\bibfnamefont {P.}~\bibnamefont
  {Masjuan}}, \bibinfo {author} {\bibfnamefont {E.}~\bibnamefont
  {Ruiz~Arriola}}, \ and\ \bibinfo {author} {\bibfnamefont {W.}~\bibnamefont
  {Broniowski}},\ }\href {\doibase 10.1103/PhysRevD.87.014005} {\bibfield
  {journal} {\bibinfo  {journal} {Phys. Rev. D}\ }\textbf {\bibinfo {volume}
  {87}},\ \bibinfo {pages} {014005} (\bibinfo {year} {2013})},\ \Eprint
  {http://arxiv.org/abs/1210.0760} {arXiv:1210.0760 [hep-ph]} \BibitemShut
  {NoStop}%
\bibitem [{\citenamefont {Fanelli}\ \emph {et~al.}(2016)\citenamefont
  {Fanelli}, \citenamefont {Pace}, \citenamefont {Romanelli}, \citenamefont
  {Salme},\ and\ \citenamefont {Salmistraro}}]{Fanelli:2016aqc}%
  \BibitemOpen
  \bibfield  {author} {\bibinfo {author} {\bibfnamefont {C.}~\bibnamefont
  {Fanelli}}, \bibinfo {author} {\bibfnamefont {E.}~\bibnamefont {Pace}},
  \bibinfo {author} {\bibfnamefont {G.}~\bibnamefont {Romanelli}}, \bibinfo
  {author} {\bibfnamefont {G.}~\bibnamefont {Salme}}, \ and\ \bibinfo {author}
  {\bibfnamefont {M.}~\bibnamefont {Salmistraro}},\ }\href {\doibase
  10.1140/epjc/s10052-016-4101-1} {\bibfield  {journal} {\bibinfo  {journal}
  {Eur. Phys. J. C}\ }\textbf {\bibinfo {volume} {76}},\ \bibinfo {pages} {253}
  (\bibinfo {year} {2016})},\ \Eprint {http://arxiv.org/abs/1603.04598}
  {arXiv:1603.04598 [hep-ph]} \BibitemShut {NoStop}%
\bibitem [{\citenamefont {Freese}\ and\ \citenamefont
  {Clo\"et}(2019)}]{Freese:2019bhb}%
  \BibitemOpen
  \bibfield  {author} {\bibinfo {author} {\bibfnamefont {A.}~\bibnamefont
  {Freese}}\ and\ \bibinfo {author} {\bibfnamefont {I.~C.}\ \bibnamefont
  {Clo\"et}},\ }\href {\doibase 10.1103/PhysRevC.100.015201} {\bibfield
  {journal} {\bibinfo  {journal} {Phys. Rev. C}\ }\textbf {\bibinfo {volume}
  {100}},\ \bibinfo {pages} {015201} (\bibinfo {year} {2019})},\ \bibinfo
  {note} {[Erratum: Phys.Rev.C 105, 059901 (2022)]},\ \Eprint
  {http://arxiv.org/abs/1903.09222} {arXiv:1903.09222 [nucl-th]} \BibitemShut
  {NoStop}%
\bibitem [{\citenamefont {Krutov}\ and\ \citenamefont
  {Troitsky}(2021)}]{Krutov:2020ewr}%
  \BibitemOpen
  \bibfield  {author} {\bibinfo {author} {\bibfnamefont {A.~F.}\ \bibnamefont
  {Krutov}}\ and\ \bibinfo {author} {\bibfnamefont {V.~E.}\ \bibnamefont
  {Troitsky}},\ }\href {\doibase 10.1103/PhysRevD.103.014029} {\bibfield
  {journal} {\bibinfo  {journal} {Phys. Rev. D}\ }\textbf {\bibinfo {volume}
  {103}},\ \bibinfo {pages} {014029} (\bibinfo {year} {2021})},\ \Eprint
  {http://arxiv.org/abs/2010.11640} {arXiv:2010.11640 [hep-ph]} \BibitemShut
  {NoStop}%
\bibitem [{\citenamefont {Xing}\ \emph {et~al.}(2023)\citenamefont {Xing},
  \citenamefont {Ding},\ and\ \citenamefont {Chang}}]{Xing:2022mvk}%
  \BibitemOpen
  \bibfield  {author} {\bibinfo {author} {\bibfnamefont {Z.}~\bibnamefont
  {Xing}}, \bibinfo {author} {\bibfnamefont {M.}~\bibnamefont {Ding}}, \ and\
  \bibinfo {author} {\bibfnamefont {L.}~\bibnamefont {Chang}},\ }\href
  {\doibase 10.1103/PhysRevD.107.L031502} {\bibfield  {journal} {\bibinfo
  {journal} {Phys. Rev. D}\ }\textbf {\bibinfo {volume} {107}},\ \bibinfo
  {pages} {L031502} (\bibinfo {year} {2023})},\ \Eprint
  {http://arxiv.org/abs/2211.06635} {arXiv:2211.06635 [hep-ph]} \BibitemShut
  {NoStop}%
\bibitem [{\citenamefont {Xu}\ \emph {et~al.}(2024)\citenamefont {Xu},
  \citenamefont {Ding}, \citenamefont {Raya}, \citenamefont {Roberts},
  \citenamefont {Rodr\'\i{}guez-Quintero},\ and\ \citenamefont
  {Schmidt}}]{Xu:2023izo}%
  \BibitemOpen
  \bibfield  {author} {\bibinfo {author} {\bibfnamefont {Y.-Z.}\ \bibnamefont
  {Xu}}, \bibinfo {author} {\bibfnamefont {M.}~\bibnamefont {Ding}}, \bibinfo
  {author} {\bibfnamefont {K.}~\bibnamefont {Raya}}, \bibinfo {author}
  {\bibfnamefont {C.~D.}\ \bibnamefont {Roberts}}, \bibinfo {author}
  {\bibfnamefont {J.}~\bibnamefont {Rodr\'\i{}guez-Quintero}}, \ and\ \bibinfo
  {author} {\bibfnamefont {S.~M.}\ \bibnamefont {Schmidt}},\ }\href {\doibase
  10.1140/epjc/s10052-024-12518-x} {\bibfield  {journal} {\bibinfo  {journal}
  {Eur. Phys. J. C}\ }\textbf {\bibinfo {volume} {84}},\ \bibinfo {pages} {191}
  (\bibinfo {year} {2024})},\ \Eprint {http://arxiv.org/abs/2311.14832}
  {arXiv:2311.14832 [hep-ph]} \BibitemShut {NoStop}%
\bibitem [{\citenamefont {Li}\ and\ \citenamefont {Vary}(2024)}]{Li:2023izn}%
  \BibitemOpen
  \bibfield  {author} {\bibinfo {author} {\bibfnamefont {Y.}~\bibnamefont
  {Li}}\ and\ \bibinfo {author} {\bibfnamefont {J.~P.}\ \bibnamefont {Vary}},\
  }\href {\doibase 10.1103/PhysRevD.109.L051501} {\bibfield  {journal}
  {\bibinfo  {journal} {Phys. Rev. D}\ }\textbf {\bibinfo {volume} {109}},\
  \bibinfo {pages} {L051501} (\bibinfo {year} {2024})},\ \Eprint
  {http://arxiv.org/abs/2312.02543} {arXiv:2312.02543 [hep-th]} \BibitemShut
  {NoStop}%
\bibitem [{\citenamefont {Liu}\ \emph {et~al.}(2024{\natexlab{a}})\citenamefont
  {Liu}, \citenamefont {Shuryak}, \citenamefont {Weiss},\ and\ \citenamefont
  {Zahed}}]{Liu:2024jno}%
  \BibitemOpen
  \bibfield  {author} {\bibinfo {author} {\bibfnamefont {W.-Y.}\ \bibnamefont
  {Liu}}, \bibinfo {author} {\bibfnamefont {E.}~\bibnamefont {Shuryak}},
  \bibinfo {author} {\bibfnamefont {C.}~\bibnamefont {Weiss}}, \ and\ \bibinfo
  {author} {\bibfnamefont {I.}~\bibnamefont {Zahed}},\ }\href {\doibase
  10.1103/PhysRevD.110.054021} {\bibfield  {journal} {\bibinfo  {journal}
  {Phys. Rev. D}\ }\textbf {\bibinfo {volume} {110}},\ \bibinfo {pages}
  {054021} (\bibinfo {year} {2024}{\natexlab{a}})},\ \Eprint
  {http://arxiv.org/abs/2405.14026} {arXiv:2405.14026 [hep-ph]} \BibitemShut
  {NoStop}%
\bibitem [{\citenamefont {Liu}\ \emph {et~al.}(2024{\natexlab{b}})\citenamefont
  {Liu}, \citenamefont {Shuryak},\ and\ \citenamefont {Zahed}}]{Liu:2024vkj}%
  \BibitemOpen
  \bibfield  {author} {\bibinfo {author} {\bibfnamefont {W.-Y.}\ \bibnamefont
  {Liu}}, \bibinfo {author} {\bibfnamefont {E.}~\bibnamefont {Shuryak}}, \ and\
  \bibinfo {author} {\bibfnamefont {I.}~\bibnamefont {Zahed}},\ }\href
  {\doibase 10.1103/PhysRevD.110.054022} {\bibfield  {journal} {\bibinfo
  {journal} {Phys. Rev. D}\ }\textbf {\bibinfo {volume} {110}},\ \bibinfo
  {pages} {054022} (\bibinfo {year} {2024}{\natexlab{b}})},\ \Eprint
  {http://arxiv.org/abs/2405.16269} {arXiv:2405.16269 [hep-ph]} \BibitemShut
  {NoStop}%
\bibitem [{\citenamefont {Wang}\ \emph
  {et~al.}(2024{\natexlab{b}})\citenamefont {Wang}, \citenamefont {Xing},
  \citenamefont {Ding}, \citenamefont {Raya},\ and\ \citenamefont
  {Chang}}]{Wang:2024sqg}%
  \BibitemOpen
  \bibfield  {author} {\bibinfo {author} {\bibfnamefont {X.}~\bibnamefont
  {Wang}}, \bibinfo {author} {\bibfnamefont {Z.}~\bibnamefont {Xing}}, \bibinfo
  {author} {\bibfnamefont {M.}~\bibnamefont {Ding}}, \bibinfo {author}
  {\bibfnamefont {K.}~\bibnamefont {Raya}}, \ and\ \bibinfo {author}
  {\bibfnamefont {L.}~\bibnamefont {Chang}},\ }\href@noop {} {\  (\bibinfo
  {year} {2024}{\natexlab{b}})},\ \Eprint {http://arxiv.org/abs/2406.09644}
  {arXiv:2406.09644 [hep-ph]} \BibitemShut {NoStop}%
\bibitem [{\citenamefont {Sultan}\ \emph {et~al.}(2024)\citenamefont {Sultan},
  \citenamefont {Xing}, \citenamefont {Raya}, \citenamefont {Bashir},\ and\
  \citenamefont {Chang}}]{Sultan:2024hep}%
  \BibitemOpen
  \bibfield  {author} {\bibinfo {author} {\bibfnamefont {M.~A.}\ \bibnamefont
  {Sultan}}, \bibinfo {author} {\bibfnamefont {Z.}~\bibnamefont {Xing}},
  \bibinfo {author} {\bibfnamefont {K.}~\bibnamefont {Raya}}, \bibinfo {author}
  {\bibfnamefont {A.}~\bibnamefont {Bashir}}, \ and\ \bibinfo {author}
  {\bibfnamefont {L.}~\bibnamefont {Chang}},\ }\href {\doibase
  10.1103/PhysRevD.110.054034} {\bibfield  {journal} {\bibinfo  {journal}
  {Phys. Rev. D}\ }\textbf {\bibinfo {volume} {110}},\ \bibinfo {pages}
  {054034} (\bibinfo {year} {2024})},\ \Eprint
  {http://arxiv.org/abs/2407.10437} {arXiv:2407.10437 [hep-ph]} \BibitemShut
  {NoStop}%
\bibitem [{\citenamefont {Fujii}\ \emph {et~al.}(2024)\citenamefont {Fujii},
  \citenamefont {Iwanaka},\ and\ \citenamefont {Tanaka}}]{Fujii:2024rqd}%
  \BibitemOpen
  \bibfield  {author} {\bibinfo {author} {\bibfnamefont {D.}~\bibnamefont
  {Fujii}}, \bibinfo {author} {\bibfnamefont {A.}~\bibnamefont {Iwanaka}}, \
  and\ \bibinfo {author} {\bibfnamefont {M.}~\bibnamefont {Tanaka}},\
  }\href@noop {} {\  (\bibinfo {year} {2024})},\ \Eprint
  {http://arxiv.org/abs/2407.21113} {arXiv:2407.21113 [hep-ph]} \BibitemShut
  {NoStop}%
\bibitem [{\citenamefont {Krutov}\ and\ \citenamefont
  {Troitsky}(2024)}]{Krutov:2024adh}%
  \BibitemOpen
  \bibfield  {author} {\bibinfo {author} {\bibfnamefont {A.~F.}\ \bibnamefont
  {Krutov}}\ and\ \bibinfo {author} {\bibfnamefont {V.~E.}\ \bibnamefont
  {Troitsky}},\ }\href@noop {} {\  (\bibinfo {year} {2024})},\ \Eprint
  {http://arxiv.org/abs/2410.17570} {arXiv:2410.17570 [hep-ph]} \BibitemShut
  {NoStop}%
\bibitem [{\citenamefont {Broniowski}\ and\ \citenamefont
  {Ruiz~Arriola}(2024)}]{Broniowski:2024oyk}%
  \BibitemOpen
  \bibfield  {author} {\bibinfo {author} {\bibfnamefont {W.}~\bibnamefont
  {Broniowski}}\ and\ \bibinfo {author} {\bibfnamefont {E.}~\bibnamefont
  {Ruiz~Arriola}},\ }\href {\doibase 10.1016/j.physletb.2024.139138} {\bibfield
   {journal} {\bibinfo  {journal} {Phys. Lett. B}\ }\textbf {\bibinfo {volume}
  {859}},\ \bibinfo {pages} {139138} (\bibinfo {year} {2024})},\ \Eprint
  {http://arxiv.org/abs/2405.07815} {arXiv:2405.07815 [hep-ph]} \BibitemShut
  {NoStop}%
\bibitem [{\citenamefont {Ruiz~Arriola}\ and\ \citenamefont
  {Broniowski}(2024)}]{RuizArriola:2024uub}%
  \BibitemOpen
  \bibfield  {author} {\bibinfo {author} {\bibfnamefont {E.}~\bibnamefont
  {Ruiz~Arriola}}\ and\ \bibinfo {author} {\bibfnamefont {W.}~\bibnamefont
  {Broniowski}}\ }(\bibinfo {year} {2024})\ \Eprint
  {http://arxiv.org/abs/2411.10354} {arXiv:2411.10354 [hep-ph]} \BibitemShut
  {NoStop}%
\bibitem [{\citenamefont {Cao}\ \emph {et~al.}(2024)\citenamefont {Cao},
  \citenamefont {Guo}, \citenamefont {Li},\ and\ \citenamefont
  {Yao}}]{Cao:2024zlf}%
  \BibitemOpen
  \bibfield  {author} {\bibinfo {author} {\bibfnamefont {X.-H.}\ \bibnamefont
  {Cao}}, \bibinfo {author} {\bibfnamefont {F.-K.}\ \bibnamefont {Guo}},
  \bibinfo {author} {\bibfnamefont {Q.-Z.}\ \bibnamefont {Li}}, \ and\ \bibinfo
  {author} {\bibfnamefont {D.-L.}\ \bibnamefont {Yao}},\ }\href@noop {} {\
  (\bibinfo {year} {2024})},\ \Eprint {http://arxiv.org/abs/2411.13398}
  {arXiv:2411.13398 [hep-ph]} \BibitemShut {NoStop}%
\bibitem [{\citenamefont {Tong}\ \emph {et~al.}(2021)\citenamefont {Tong},
  \citenamefont {Ma},\ and\ \citenamefont {Yuan}}]{Tong:2021ctu}%
  \BibitemOpen
  \bibfield  {author} {\bibinfo {author} {\bibfnamefont {X.-B.}\ \bibnamefont
  {Tong}}, \bibinfo {author} {\bibfnamefont {J.-P.}\ \bibnamefont {Ma}}, \ and\
  \bibinfo {author} {\bibfnamefont {F.}~\bibnamefont {Yuan}},\ }\href {\doibase
  10.1016/j.physletb.2021.136751} {\bibfield  {journal} {\bibinfo  {journal}
  {Phys. Lett. B}\ }\textbf {\bibinfo {volume} {823}},\ \bibinfo {pages}
  {136751} (\bibinfo {year} {2021})},\ \Eprint
  {http://arxiv.org/abs/2101.02395} {arXiv:2101.02395 [hep-ph]} \BibitemShut
  {NoStop}%
\bibitem [{\citenamefont {Tong}\ \emph {et~al.}(2022)\citenamefont {Tong},
  \citenamefont {Ma},\ and\ \citenamefont {Yuan}}]{Tong:2022zax}%
  \BibitemOpen
  \bibfield  {author} {\bibinfo {author} {\bibfnamefont {X.-B.}\ \bibnamefont
  {Tong}}, \bibinfo {author} {\bibfnamefont {J.-P.}\ \bibnamefont {Ma}}, \ and\
  \bibinfo {author} {\bibfnamefont {F.}~\bibnamefont {Yuan}},\ }\href {\doibase
  10.1007/JHEP10(2022)046} {\bibfield  {journal} {\bibinfo  {journal} {JHEP}\
  }\textbf {\bibinfo {volume} {10}},\ \bibinfo {pages} {046} (\bibinfo {year}
  {2022})},\ \Eprint {http://arxiv.org/abs/2203.13493} {arXiv:2203.13493
  [hep-ph]} \BibitemShut {NoStop}%
\bibitem [{\citenamefont {Pokorski}(2005)}]{Pokorski:1987ed}%
  \BibitemOpen
  \bibfield  {author} {\bibinfo {author} {\bibfnamefont {S.}~\bibnamefont
  {Pokorski}},\ }\href@noop {} {\emph {\bibinfo {title} {{\it Gauge field
  theories}}}}\ (\bibinfo  {publisher} {Cambridge University Press},\ \bibinfo
  {year} {2005})\BibitemShut {NoStop}%
\bibitem [{\citenamefont {Belitsky}\ and\ \citenamefont
  {Radyushkin}(2005)}]{Belitsky:2005qn}%
  \BibitemOpen
  \bibfield  {author} {\bibinfo {author} {\bibfnamefont {A.~V.}\ \bibnamefont
  {Belitsky}}\ and\ \bibinfo {author} {\bibfnamefont {A.~V.}\ \bibnamefont
  {Radyushkin}},\ }\href {\doibase 10.1016/j.physrep.2005.06.002} {\bibfield
  {journal} {\bibinfo  {journal} {Phys. Rept.}\ }\textbf {\bibinfo {volume}
  {418}},\ \bibinfo {pages} {1} (\bibinfo {year} {2005})},\ \Eprint
  {http://arxiv.org/abs/hep-ph/0504030} {arXiv:hep-ph/0504030} \BibitemShut
  {NoStop}%
\bibitem [{\citenamefont {Soper}(1977)}]{PhysRevD.15.1141}%
  \BibitemOpen
  \bibfield  {author} {\bibinfo {author} {\bibfnamefont {D.~E.}\ \bibnamefont
  {Soper}},\ }\href {\doibase 10.1103/PhysRevD.15.1141} {\bibfield  {journal}
  {\bibinfo  {journal} {Phys. Rev. D}\ }\textbf {\bibinfo {volume} {15}},\
  \bibinfo {pages} {1141} (\bibinfo {year} {1977})}\BibitemShut {NoStop}%
\bibitem [{\citenamefont {Burkardt}(2000)}]{Burkardt:2000za}%
  \BibitemOpen
  \bibfield  {author} {\bibinfo {author} {\bibfnamefont {M.}~\bibnamefont
  {Burkardt}},\ }\href {\doibase 10.1103/PhysRevD.62.071503} {\bibfield
  {journal} {\bibinfo  {journal} {Phys. Rev. D}\ }\textbf {\bibinfo {volume}
  {62}},\ \bibinfo {pages} {071503} (\bibinfo {year} {2000})},\ \bibinfo {note}
  {[Erratum: Phys.Rev.D 66, 119903 (2002)]},\ \Eprint
  {http://arxiv.org/abs/hep-ph/0005108} {arXiv:hep-ph/0005108} \BibitemShut
  {NoStop}%
\bibitem [{\citenamefont {Diehl}(2002)}]{Diehl:2002he}%
  \BibitemOpen
  \bibfield  {author} {\bibinfo {author} {\bibfnamefont {M.}~\bibnamefont
  {Diehl}},\ }\href {\doibase 10.1007/s10052-002-1016-9} {\bibfield  {journal}
  {\bibinfo  {journal} {Eur. Phys. J. C}\ }\textbf {\bibinfo {volume} {25}},\
  \bibinfo {pages} {223} (\bibinfo {year} {2002})},\ \bibinfo {note} {[Erratum:
  Eur.Phys.J.C 31, 277--278 (2003)]},\ \Eprint
  {http://arxiv.org/abs/hep-ph/0205208} {arXiv:hep-ph/0205208} \BibitemShut
  {NoStop}%
\bibitem [{\citenamefont {Burkardt}(2003)}]{Burkardt:2002hr}%
  \BibitemOpen
  \bibfield  {author} {\bibinfo {author} {\bibfnamefont {M.}~\bibnamefont
  {Burkardt}},\ }\href {\doibase 10.1142/S0217751X03012370} {\bibfield
  {journal} {\bibinfo  {journal} {Int. J. Mod. Phys. A}\ }\textbf {\bibinfo
  {volume} {18}},\ \bibinfo {pages} {173} (\bibinfo {year} {2003})},\ \Eprint
  {http://arxiv.org/abs/hep-ph/0207047} {arXiv:hep-ph/0207047} \BibitemShut
  {NoStop}%
\bibitem [{\citenamefont {Miller}(2010)}]{Miller:2010nz}%
  \BibitemOpen
  \bibfield  {author} {\bibinfo {author} {\bibfnamefont {G.~A.}\ \bibnamefont
  {Miller}},\ }\href {\doibase 10.1146/annurev.nucl.012809.104508} {\bibfield
  {journal} {\bibinfo  {journal} {Ann. Rev. Nucl. Part. Sci.}\ }\textbf
  {\bibinfo {volume} {60}},\ \bibinfo {pages} {1} (\bibinfo {year} {2010})},\
  \Eprint {http://arxiv.org/abs/1002.0355} {arXiv:1002.0355 [nucl-th]}
  \BibitemShut {NoStop}%
\bibitem [{\citenamefont {Freese}\ and\ \citenamefont
  {Miller}(2023)}]{Freese:2022fat}%
  \BibitemOpen
  \bibfield  {author} {\bibinfo {author} {\bibfnamefont {A.}~\bibnamefont
  {Freese}}\ and\ \bibinfo {author} {\bibfnamefont {G.~A.}\ \bibnamefont
  {Miller}},\ }\href {\doibase 10.1103/PhysRevD.108.034008} {\bibfield
  {journal} {\bibinfo  {journal} {Phys. Rev. D}\ }\textbf {\bibinfo {volume}
  {108}},\ \bibinfo {pages} {034008} (\bibinfo {year} {2023})},\ \Eprint
  {http://arxiv.org/abs/2210.03807} {arXiv:2210.03807 [hep-ph]} \BibitemShut
  {NoStop}%
\bibitem [{\citenamefont {Pobylitsa}(2002)}]{Pobylitsa:2002iu}%
  \BibitemOpen
  \bibfield  {author} {\bibinfo {author} {\bibfnamefont {P.~V.}\ \bibnamefont
  {Pobylitsa}},\ }\href@noop {} {\bibfield  {journal} {\bibinfo  {journal}
  {Phys. Rev.}\ }\textbf {\bibinfo {volume} {D66}},\ \bibinfo {pages} {094002}
  (\bibinfo {year} {2002})},\ \Eprint {http://arxiv.org/abs/hep-ph/0204337}
  {hep-ph/0204337} \BibitemShut {NoStop}%
\bibitem [{\citenamefont {Freese}\ and\ \citenamefont
  {Miller}(2022)}]{Freese:2021mzg}%
  \BibitemOpen
  \bibfield  {author} {\bibinfo {author} {\bibfnamefont {A.}~\bibnamefont
  {Freese}}\ and\ \bibinfo {author} {\bibfnamefont {G.~A.}\ \bibnamefont
  {Miller}},\ }\href {\doibase 10.1103/PhysRevD.105.014003} {\bibfield
  {journal} {\bibinfo  {journal} {Phys. Rev. D}\ }\textbf {\bibinfo {volume}
  {105}},\ \bibinfo {pages} {014003} (\bibinfo {year} {2022})},\ \Eprint
  {http://arxiv.org/abs/2108.03301} {arXiv:2108.03301 [hep-ph]} \BibitemShut
  {NoStop}%
\bibitem [{\citenamefont {Panteleeva}\ and\ \citenamefont
  {Polyakov}(2021)}]{Panteleeva:2021iip}%
  \BibitemOpen
  \bibfield  {author} {\bibinfo {author} {\bibfnamefont {J.~Y.}\ \bibnamefont
  {Panteleeva}}\ and\ \bibinfo {author} {\bibfnamefont {M.~V.}\ \bibnamefont
  {Polyakov}},\ }\href {\doibase 10.1103/PhysRevD.104.014008} {\bibfield
  {journal} {\bibinfo  {journal} {Phys. Rev. D}\ }\textbf {\bibinfo {volume}
  {104}},\ \bibinfo {pages} {014008} (\bibinfo {year} {2021})},\ \Eprint
  {http://arxiv.org/abs/2102.10902} {arXiv:2102.10902 [hep-ph]} \BibitemShut
  {NoStop}%
\bibitem [{\citenamefont {Donoghue}\ and\ \citenamefont
  {Na}(1997)}]{Donoghue:1996bt}%
  \BibitemOpen
  \bibfield  {author} {\bibinfo {author} {\bibfnamefont {J.~F.}\ \bibnamefont
  {Donoghue}}\ and\ \bibinfo {author} {\bibfnamefont {E.~S.}\ \bibnamefont
  {Na}},\ }\href {\doibase 10.1103/PhysRevD.56.7073} {\bibfield  {journal}
  {\bibinfo  {journal} {Phys. Rev. D}\ }\textbf {\bibinfo {volume} {56}},\
  \bibinfo {pages} {7073} (\bibinfo {year} {1997})},\ \Eprint
  {http://arxiv.org/abs/hep-ph/9611418} {arXiv:hep-ph/9611418} \BibitemShut
  {NoStop}%
\bibitem [{\citenamefont {Gasser}\ and\ \citenamefont
  {Leutwyler}(1984)}]{Gasser:1983yg}%
  \BibitemOpen
  \bibfield  {author} {\bibinfo {author} {\bibfnamefont {J.}~\bibnamefont
  {Gasser}}\ and\ \bibinfo {author} {\bibfnamefont {H.}~\bibnamefont
  {Leutwyler}},\ }\href {\doibase 10.1016/0003-4916(84)90242-2} {\bibfield
  {journal} {\bibinfo  {journal} {Annals Phys.}\ }\textbf {\bibinfo {volume}
  {158}},\ \bibinfo {pages} {142} (\bibinfo {year} {1984})}\BibitemShut
  {NoStop}%
\bibitem [{\citenamefont {Miller}\ \emph
  {et~al.}(2011{\natexlab{a}})\citenamefont {Miller}, \citenamefont
  {Strikman},\ and\ \citenamefont {Weiss}}]{Miller:2010tz}%
  \BibitemOpen
  \bibfield  {author} {\bibinfo {author} {\bibfnamefont {G.~A.}\ \bibnamefont
  {Miller}}, \bibinfo {author} {\bibfnamefont {M.}~\bibnamefont {Strikman}}, \
  and\ \bibinfo {author} {\bibfnamefont {C.}~\bibnamefont {Weiss}},\ }\href
  {\doibase 10.1103/PhysRevD.83.013006} {\bibfield  {journal} {\bibinfo
  {journal} {Phys. Rev. D}\ }\textbf {\bibinfo {volume} {83}},\ \bibinfo
  {pages} {013006} (\bibinfo {year} {2011}{\natexlab{a}})},\ \Eprint
  {http://arxiv.org/abs/1011.1472} {arXiv:1011.1472 [hep-ph]} \BibitemShut
  {NoStop}%
\bibitem [{\citenamefont {Miller}\ \emph
  {et~al.}(2011{\natexlab{b}})\citenamefont {Miller}, \citenamefont
  {Strikman},\ and\ \citenamefont {Weiss}}]{Miller:2011du}%
  \BibitemOpen
  \bibfield  {author} {\bibinfo {author} {\bibfnamefont {G.~A.}\ \bibnamefont
  {Miller}}, \bibinfo {author} {\bibfnamefont {M.}~\bibnamefont {Strikman}}, \
  and\ \bibinfo {author} {\bibfnamefont {C.}~\bibnamefont {Weiss}},\ }\href
  {\doibase 10.1103/PhysRevC.84.045205} {\bibfield  {journal} {\bibinfo
  {journal} {Phys. Rev. C}\ }\textbf {\bibinfo {volume} {84}},\ \bibinfo
  {pages} {045205} (\bibinfo {year} {2011}{\natexlab{b}})},\ \Eprint
  {http://arxiv.org/abs/1105.6364} {arXiv:1105.6364 [hep-ph]} \BibitemShut
  {NoStop}%
\bibitem [{\citenamefont {Ruiz~Arriola}\ and\ \citenamefont
  {Sanchez-Puertas}(2024)}]{RuizArriola:2024gwb}%
  \BibitemOpen
  \bibfield  {author} {\bibinfo {author} {\bibfnamefont {E.}~\bibnamefont
  {Ruiz~Arriola}}\ and\ \bibinfo {author} {\bibfnamefont {P.}~\bibnamefont
  {Sanchez-Puertas}},\ }\href {\doibase 10.1103/PhysRevD.110.054003} {\bibfield
   {journal} {\bibinfo  {journal} {Phys. Rev. D}\ }\textbf {\bibinfo {volume}
  {110}},\ \bibinfo {pages} {054003} (\bibinfo {year} {2024})},\ \Eprint
  {http://arxiv.org/abs/2403.07121} {arXiv:2403.07121 [hep-ph]} \BibitemShut
  {NoStop}%
\bibitem [{\citenamefont {Sanchez-Puertas}\ and\ \citenamefont
  {Ruiz~Arriola}(2024)}]{Sanchez-Puertas:2024siv}%
  \BibitemOpen
  \bibfield  {author} {\bibinfo {author} {\bibfnamefont {P.}~\bibnamefont
  {Sanchez-Puertas}}\ and\ \bibinfo {author} {\bibfnamefont {E.}~\bibnamefont
  {Ruiz~Arriola}},\ }in\ \href@noop {} {\emph {\bibinfo {booktitle} {{10th
  International Conference on Quarks and Nuclear Physics}}}}\ (\bibinfo {year}
  {2024})\ \Eprint {http://arxiv.org/abs/2410.17804} {arXiv:2410.17804
  [hep-ph]} \BibitemShut {NoStop}%
\bibitem [{\citenamefont {Ruiz~Arriola}\ \emph {et~al.}(2024)\citenamefont
  {Ruiz~Arriola}, \citenamefont {Sanchez-Puertas},\ and\ \citenamefont
  {Weiss}}]{RuizArriola:2024}%
  \BibitemOpen
  \bibfield  {author} {\bibinfo {author} {\bibfnamefont {E.}~\bibnamefont
  {Ruiz~Arriola}}, \bibinfo {author} {\bibfnamefont {P.}~\bibnamefont
  {Sanchez-Puertas}}, \ and\ \bibinfo {author} {\bibfnamefont {C.}~\bibnamefont
  {Weiss}},\ }\href@noop {} {\bibfield  {journal} {\bibinfo  {journal} {Work in
  preparation}\ } (\bibinfo {year} {2024})}\BibitemShut {NoStop}%
\bibitem [{\citenamefont {Miller}(2009)}]{Miller:2009qu}%
  \BibitemOpen
  \bibfield  {author} {\bibinfo {author} {\bibfnamefont {G.~A.}\ \bibnamefont
  {Miller}},\ }\href {\doibase 10.1103/PhysRevC.79.055204} {\bibfield
  {journal} {\bibinfo  {journal} {Phys. Rev. C}\ }\textbf {\bibinfo {volume}
  {79}},\ \bibinfo {pages} {055204} (\bibinfo {year} {2009})},\ \Eprint
  {http://arxiv.org/abs/0901.1117} {arXiv:0901.1117 [nucl-th]} \BibitemShut
  {NoStop}%
\bibitem [{\citenamefont {Colangelo}\ \emph {et~al.}(2001)\citenamefont
  {Colangelo}, \citenamefont {Gasser},\ and\ \citenamefont
  {Leutwyler}}]{Colangelo:2001df}%
  \BibitemOpen
  \bibfield  {author} {\bibinfo {author} {\bibfnamefont {G.}~\bibnamefont
  {Colangelo}}, \bibinfo {author} {\bibfnamefont {J.}~\bibnamefont {Gasser}}, \
  and\ \bibinfo {author} {\bibfnamefont {H.}~\bibnamefont {Leutwyler}},\ }\href
  {\doibase 10.1016/S0550-3213(01)00147-X} {\bibfield  {journal} {\bibinfo
  {journal} {Nucl. Phys. B}\ }\textbf {\bibinfo {volume} {603}},\ \bibinfo
  {pages} {125} (\bibinfo {year} {2001})},\ \Eprint
  {http://arxiv.org/abs/hep-ph/0103088} {arXiv:hep-ph/0103088} \BibitemShut
  {NoStop}%
\bibitem [{\citenamefont {Garcia-Martin}\ \emph {et~al.}(2011)\citenamefont
  {Garcia-Martin}, \citenamefont {Kaminski}, \citenamefont {Pelaez},
  \citenamefont {Ruiz~de Elvira},\ and\ \citenamefont
  {Yndurain}}]{Garcia-Martin:2011iqs}%
  \BibitemOpen
  \bibfield  {author} {\bibinfo {author} {\bibfnamefont {R.}~\bibnamefont
  {Garcia-Martin}}, \bibinfo {author} {\bibfnamefont {R.}~\bibnamefont
  {Kaminski}}, \bibinfo {author} {\bibfnamefont {J.~R.}\ \bibnamefont
  {Pelaez}}, \bibinfo {author} {\bibfnamefont {J.}~\bibnamefont {Ruiz~de
  Elvira}}, \ and\ \bibinfo {author} {\bibfnamefont {F.~J.}\ \bibnamefont
  {Yndurain}},\ }\href {\doibase 10.1103/PhysRevD.83.074004} {\bibfield
  {journal} {\bibinfo  {journal} {Phys. Rev. D}\ }\textbf {\bibinfo {volume}
  {83}},\ \bibinfo {pages} {074004} (\bibinfo {year} {2011})},\ \Eprint
  {http://arxiv.org/abs/1102.2183} {arXiv:1102.2183 [hep-ph]} \BibitemShut
  {NoStop}%
\bibitem [{\citenamefont {Ruiz~Arriola}\ and\ \citenamefont
  {Broniowski}(2008)}]{RuizArriola:2008sq}%
  \BibitemOpen
  \bibfield  {author} {\bibinfo {author} {\bibfnamefont {E.}~\bibnamefont
  {Ruiz~Arriola}}\ and\ \bibinfo {author} {\bibfnamefont {W.}~\bibnamefont
  {Broniowski}},\ }\href {\doibase 10.1103/PhysRevD.78.034031} {\bibfield
  {journal} {\bibinfo  {journal} {Phys. Rev. D}\ }\textbf {\bibinfo {volume}
  {78}},\ \bibinfo {pages} {034031} (\bibinfo {year} {2008})},\ \Eprint
  {http://arxiv.org/abs/0807.3488} {arXiv:0807.3488 [hep-ph]} \BibitemShut
  {NoStop}%
\end{thebibliography}%

\end{document}